\documentclass[12pt,a4paper]{article}
\pdfoutput=1

\usepackage[T1]{fontenc} 
\usepackage{amsmath}
\usepackage{amsfonts}
\usepackage{amssymb}
\usepackage{graphicx}
\usepackage{subfig}
\usepackage{hyperref}
\usepackage{color}

\usepackage{fancyhdr}
\usepackage{indentfirst}	
\usepackage{amsmath,amssymb,latexsym,epsf,epsfig,graphicx,tikz}
\usepackage[utf8]{inputenc}
\usepackage{empheq}
\usepackage{hyperref}
\usepackage{amsthm}
\usepackage{eucal}
\usepackage{eufrak}
\usepackage{mathtools}
\usepackage{mathrsfs}
\usepackage{array}
\usepackage[makeroom]{cancel}
\usepackage{verbatim}

\newcommand{\D}{{\rm d}}

\newcommand{\upleft}[2]{\prescript{(#1)}{}{\! #2}}

\newcommand{\beq}{\begin{equation}}
\newcommand{\eeq}{\end{equation}}
\newcommand{\bea}{\begin{eqnarray}}
\newcommand{\eea}{\end{eqnarray}}
\def\be{\begin{equation}}
\def\ee{\end{equation}}

\def\beq{\begin{equation}}
\def\eeq{\end{equation}}

\newcommand{\Mpl}{M_{\textrm{Pl}}}

\def\({\left(}
\def\){\right)}

\def\mpl{M_{\rm Pl}}
\def\p{\partial}

\def\lsim{\mathrel{\rlap{\lower3pt\hbox{\hskip0pt$\sim$}}
     \raise1pt\hbox{$<$}}}         
\def\gsim{\mathrel{\rlap{\lower4pt\hbox{\hskip1pt$\sim$}}
     \raise1pt\hbox{$>$}}}         
\def\lsim{\mathrel{\rlap{\lower3pt\hbox{\hskip0pt$\sim$}}
     \raise1pt\hbox{$<$}}}         
\def\gsim{\mathrel{\rlap{\lower4pt\hbox{\hskip1pt$\sim$}}
     \raise1pt\hbox{$>$}}}         
\def\beq{\begin{eqnarray}}
\def\eeq{\end{eqnarray}}
\def\ba{\begin{eqnarray}}
\def\ea{\end{eqnarray}}
\def\({\left(}
\def\){\right)}

\begin{document}
\def\thefootnote{\fnsymbol{footnote}}

\begin{center}
\LARGE{{\bf Constraints on Single-Field Inflation}}\\

\end{center}
\begin{center}
\vspace{.3cm}

\large{David Pirtskhalava$^{\rm a}$, Luca Santoni$^{\rm b,c}$, Enrico Trincherini$^{\rm b,c}$}
\\[0.7cm]

\normalsize{
\textit{$^a$Institut de Th\'eorie des Ph\'enom\'enes Physiques\\
EPFL Lausanne, Switzerland}}
\vspace{.2cm}

\normalsize{
\textit{$^b$Scuola Normale Superiore, Piazza dei Cavalieri 7, 56126, Pisa, Italy}}

\vspace{.2cm}

\normalsize{
\textit{$^c$INFN -- Sezione di Pisa, 56200, Pisa, Italy}}

\vspace{.2cm}

\vspace{.8cm}

\hrule \vspace{0.3cm}
\noindent \small{\textbf{Abstract}\\
} 
\end{center}
Many alternatives to canonical slow-roll inflation have been proposed over the years, one of the main motivations being to have a model, capable of generating observable values of non-Gaussianity. In this work, we (re-)explore the physical implications of a great majority of such models within a single, effective field theory framework (including novel models with large non-Gaussianity discussed for the first time below.) The constraints we apply---both theoretical and experimental---are found to be rather robust,  
determined to a great extent by just three parameters: the coefficients of the 
quadratic EFT operators $(\delta N)^2$ and $\delta N \delta E$, and the slow-roll parameter $\varepsilon$.
This allows to significantly limit the majority of single-field alternatives to canonical slow-roll inflation. While the existing data still leaves some room for most of the considered models, the situation would change dramatically if the current upper limit on the tensor-to-scalar ratio decreased down to $r <10^{-2}$. Apart from inflationary models driven by plateau-like potentials, the single-field model that would have a chance of surviving this bound is the recently proposed slow-roll inflation with weakly-broken galileon symmetry. In contrast to \textit{canonical} slow-roll inflation, the latter model can support $r<10^{-2}$ even if driven by a convex potential, as well as generate observable values for the amplitude of non-Gaussianity. 
\\ 
\noindent
\hrule
\def\thefootnote{\arabic{footnote}}
\setcounter{footnote}{0}

\newpage
\section{Introduction and summary}

In the last decades cosmology has seen a remarkable transformation into a precision science. The ongoing experimental program aims at constraining the properties of the universe to an unprecedented accuracy, the ultimate goal being to shed light on the precise physics governing its dynamics. 

Inflation provides perhaps the most compelling picture of the universe at the early stages of its history. Although extremely successful as a paradigm, it is fair to say that the details of the microscopic physics behind it are still far from clear: there are many models, and a certain fraction of these is capable of reproducing the current observational data quite well. While awaiting further experimental input, it is thus important to look for ways of prioritizing the existing list of inflationary scenarios. 

A particularly convenient framework for carrying out model-independent analyses of single field inflation has been suggested in \cite{Creminelli:2006xe,Cheung:2007st}, and goes under the name of the \textit{effective field theory of inflation} (EFTI) (see Ref. \cite{Weinberg:2008hq} for a complementary approach to the problem.) The relevant effective theory is formulated in the \textit{unitary gauge}, in which the inflation perturbations are frozen (or, in other words, `eaten' by the metric), $\phi(x,t)=\phi_0(t)$. One then writes down all possible operators in derivative expansion, consistent with the unbroken symmetries; for single-field inflation, the latter are arbitrary reparametrizations of spatial coordinates, $x^i\to x^i+\xi^i(x,t)$, and the most general unitary-gauge action reads \cite{Cheung:2007st}
\begin{equation}
\begin{split}
S &= \int\D^4x\sqrt{-g}\,
	\bigg[\frac{M_P^2}{2}\left(\upleft{3}{R}+\frac{E_{ij}E^{ij}-E^2}{N^2}\right) 
	- \frac{M_P^2\dot{H}}{N^2} - M_P^2(3H^2+\dot{H}) +
\\
&\qquad
	+ \frac{M_2^4}{2}(\delta N)^2 + M_3^4(\delta N)^3 - \hat{M}_1^3\delta N \delta E + \hat{M}_2^3(\delta N)^2 \delta E + ... \bigg] ~,
\label{action}
\end{split}
\end{equation}
where by the ellipsis we have denoted terms of higher order in field perturbations\footnote{By `$\delta$' we denote the perturbation of the relevant quantity over its background value. In what follows, we will be interested in the effective action up to the 3rd order in the field perturbations.} and/or spacetime derivatives. The building blocks of the relevant effective field theory are quantities covariant under the three-dimensional diffeomorphisms, such as the lapse variable $N$, the intrinsic and extrinsic curvatures of equal-time hypersurfaces $\upleft{3}{R}$ and $K_{ij}= E_{ij}/N$, etc., see Appendix \ref{appa} for a detailed account. While the first line of \eqref{action} is fixed by the background equations of motion, the coefficients of the operators in the second line are unconstrained and parametrize all possible single-field models on a quasi de-Sitter space with a Hubble rate $H$ \cite{Cheung:2007st}. 

The free parameters of the effective theory \eqref{action} depend on the underlying model of inflation. For canonical slow-roll inflation, for example, they all vanish. More generally, in the spirit of EFT, one expects that the physical observables are predominantly determined by operators with the least number of space/time derivatives. In particular, if the unitary-gauge action \eqref{action} stems from an effective field theory of the inflaton with a single characteristic cutoff scale $\Lambda$ and with no unnatural parameters, one can show that
the leading deviations from slow-roll inflation are determined by the coefficients $M_2^4, ~M_3^4$, etc. of the operators with no derivatives. The rest of the operators give rise to effects further suppressed by powers of the small ratio $\Lambda/\mpl$, and the derivative expansion applies in the unitary gauge in a straightforward way. 

The effective field theories of the type \eqref{action} with non-zero coefficients $M_2^4$ and $M_3^4$ arise from models such as $k-$inflation \cite{ArmendarizPicon:1999rj} and DBI inflation \cite{Silverstein:2003hf,Alishahiha:2004eh} (strictly speaking, it is only the latter model that can be considered a well-defined EFT in terms of the original inflaton field.) The experimental constraints on DBI-like theories, described by the action \eqref{action} with $\hat{M}_1^3=\hat{M}_2^3=0$ are well-known \cite{Senatore:2009gt,D'Amico:2014cya}. A particularly interesting region of the parameter space is the one corresponding to strongly subluminal scalar perturbations, $M_2^4 \gg \mpl^2 |\dot H|$, in which case there is the possibility to generate sizeable non-Gaussianity within the regime of validity of the low-energy effective theory. The existing experimental data still leaves room for detecting interesting deviations from canonical slow-roll inflation within the DBI framework, and indeed, any possible hints of (equilateral) non-Gaussianity would make a strong case for these models. 

In this paper we wish to explore the status of a more general effective theory of inflation, characterized by non-zero $\hat{M}_1^3$ and $\hat{M}_2^3$. At first sight, the above remarks concerning the derivative expansion imply that large effects from the corresponding operators should not be anticipated: they are higher-derivative, and are therefore expected to play a minor role. There are however cases where the effects associated with these operators can, and do, become large. This happens in theories characterized by \textit{weakly broken} \cite{Pirtskhalava:2015nla} \textit{invariance under galileon transformations} \cite{Nicolis:2008in}, which in the unitary gauge take on the following form
\beq
\label{gi}
t\to t+ b_\mu x^\mu~.
\eeq
Whenever there exists an (approximate) invariance under \eqref{gi}, the magnitudes of the EFT coefficients in the action \eqref{action} are governed by the following \textit{radiatively stable} hierarchy \cite{Pirtskhalava:2015nla}
\beq
\label{hierarchy}
\big |M_2^4\big | \sim \big | M_3^4\big |\sim \big |\hat{M}_1^3 H\big |\sim \big |\hat{M}_2^3 H\big | ~.
\eeq
It is then straightforward to show that all of the operators in \eqref{action} play an \textit{equally important role} for the dynamics of perturbations. 
Furthermore, even for the simplest theories with weakly broken galileon (WBG) symmetry, the coefficients in \eqref{hierarchy} are generically independent of each other.

\begin{figure}
\centering
\includegraphics[width=.75\textwidth]{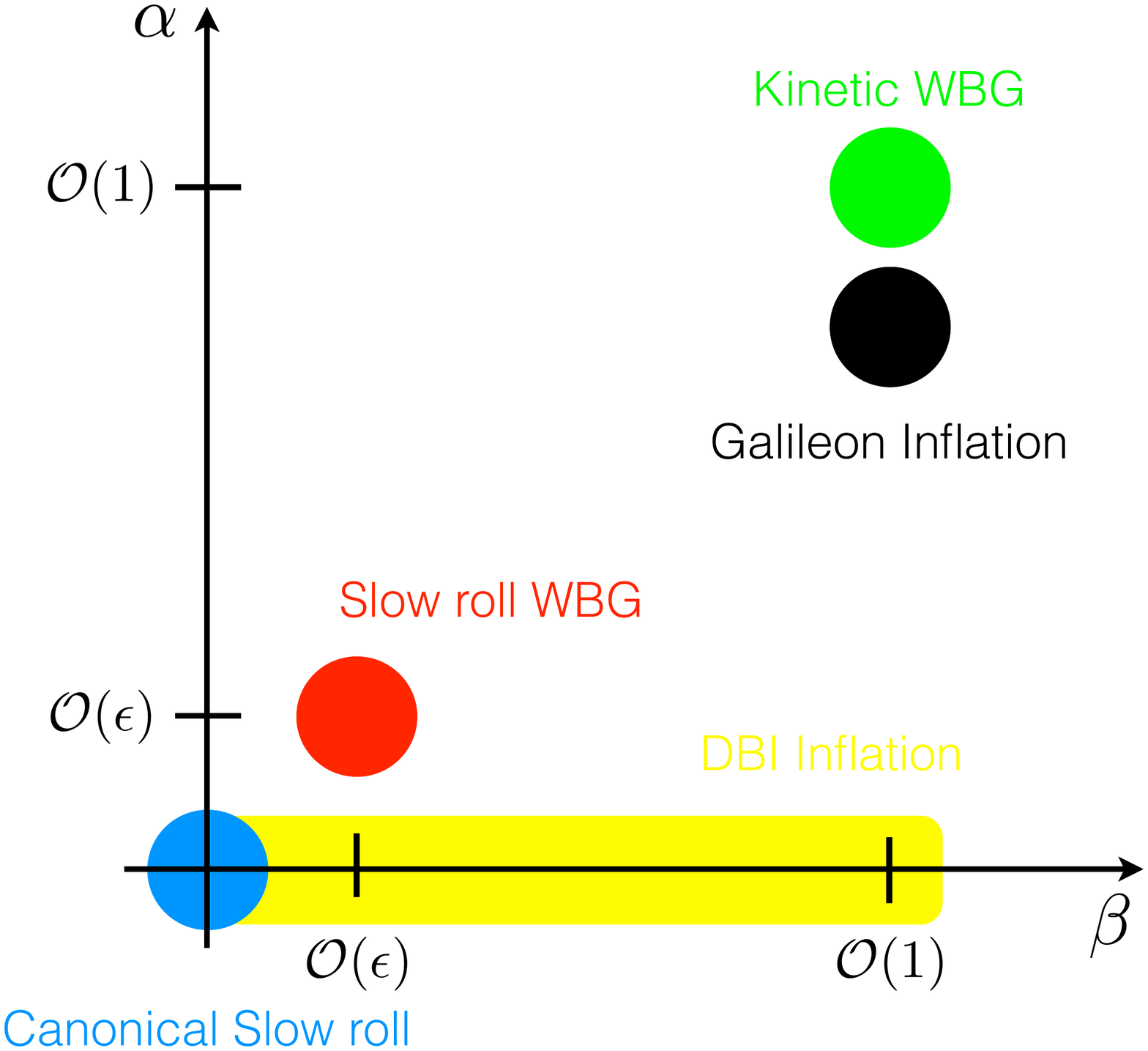} 
\caption{Various single-field models of inflation in the $\alpha$--$\beta$ plane.}
\label{fig:models}
\end{figure}

In order to proceed, one needs a guideline concerning magnitudes of the above EFT coefficients that can arise from a sensible inflationary model.
In fact, there are good reasons to believe that the coefficients in \eqref{hierarchy} can not parametrically exceed $\mpl^2 H^2$ in magnitude if the inflationary background under consideration is to be insensitive to loop corrections \footnote{The theories with WBG symmetry can saturate this upper bound within a well-defined low-energy EFT.} \cite{Pirtskhalava:2015nla}. This motivates the following parametrization of the unitary-gauge effective theory \eqref{action}
\beq
\label{coeffs}
\alpha =\frac{\hat M_1^3}{2\mpl^2 H}~, \quad \beta =\frac{M_2^4}{2\mpl^2 H^2}~,\quad \gamma=\frac{\hat M_2^3}{\mpl^2 H}~, \quad \delta = \frac{M_3^4}{\mpl^2 H^2}~.
\eeq
The values of the above dimensionless coefficients will encode the deviations of a given model from canonical slow-roll inflation, as well as distinguish between various single-field models. The cartoon in Fig. \ref{fig:models} illustrates the regions in the $\alpha$--$\beta$ plane, occupied by the theories that we consider in what follows.

With the above remarks in mind, it is of some interest to study the observational status of the general parameter space defined by\footnote{To be as general as possible in the analysis below we will abuse our theoretical expectations by allowing a slightly larger---order ten---upper bound on the dimensionless EFT parameters.}
\beq
\label{ps}
0\lsim \big \{~|\alpha|, ~|\beta|, ~|\gamma|,~ |\delta| ~\big \} \lsim \mathcal{O}\(1\)~.
\eeq
This parameter space, as we argue in the next section, captures a great majority of existing single-field inflationary models and one purpose of the present paper is precisely to explore it in the light of the present experimental results.

As a technical, but crucial remark, we note that for the most general values of the parameters from the range \eqref{ps}, the so-called \textit{decoupling limit} (DL) analysis of perturbations is not available. Indeed, the DL focuses on the dynamics of the Goldstone boson of time translation symmetry breaking $\pi$, ignoring its mixings with the metric degrees of freedom (this takes advantage of what is a direct analog of the \textit{Goldstone boson equivalence theorem} in massive spin-1 theories \cite{Lee:1977eg}.) While in many cases this is an adequate approximation at the relevant energy scales (i.e. those of order the inflationary Hubble rate), we will see that close to saturating the upper bound in \eqref{ps}, mixings of $\pi$ with gravity become order-one important -- invalidating any DL calculation of the scalar perturbations' properties. Since we will find that various theoretical/experimental constraints already cover much of the parameter space of interest, even order-one theoretical effects can be important for studying the viability of some of the theories we consider below. 

Our calculation of the full cubic scalar action in the EFT \eqref{action} comes with several interesting by-products. Namely, we find that for a certain part of the parameter space, not only is mixing with gravity non-negligible, but it turns out to completely dominate non-Gaussianity, giving rise to the relation 
\beq
\label{largeng}
f_{\rm NL}\propto \frac{1}{c_s^6}~
\eeq
in the limit of a small speed of sound $c_s$~.
Such an abrupt growth of $f_{\rm NL}$ would have never been seen from a decoupling limit perspective, which gives $f_{\rm NL}\propto 1/c_s^2$ for the same region of the parameter space\footnote{See Refs. \cite{Mizuno:2010ag,Burrage:2010cu} for the DL treatment.} (the regime characterized by \eqref{largeng} is studied in great detail in Appendix \ref{appc}~.) 
As interesting as it is, however, we unfortunately find that \eqref{largeng} can only be of academic interest: the parameter space giving rise to such values of $f_{\rm NL}$ is already ruled out by the existing experimental bounds on primordial gravitational waves.

Another interesting effect is the inverse-proportional growth (for non-zero $\gamma$ and $\delta$) of the amplitude of non-Gaussianity with a small tensor-to-scalar ratio $r$~:
\beq
\label{1overr}
f_{\rm NL} = \gamma~\frac{c_s}{r}~ \frac{80}{81}~\frac{\alpha^2-3\alpha+2-\varepsilon+3(\alpha-1)^2c_s^2}{ (\alpha-1)^4} -\delta ~\frac{c_s^3}{r} ~\frac{80}{81}~\frac{1}{ (\alpha-1)^3} + \dots~,
\eeq
where we have defined $\varepsilon\equiv -\dot H/H^2$~. This formula, supplemented by the present limits on the primordial gravitational waves, will play an important role in constraining the theories under consideration.

As to the phenomenology, we find that the constraints on the effective theory \eqref{action} are rather robust: despite the apparent multitude of the EFT coefficients, basic theoretical considerations (such as the absence of instabilities and of superluminal scalar modes) and current limits on the primordial gravitational waves and on non-Gaussianity already limit most of the parameter space. These constraints operate in a coordinated way, ruling out complementary regions of the latter. Moreover, we find that the theoretical and experimental viability of a given region of the parameter space is to a great extent determined by the set of just three numbers characterizing the operators in \eqref{action}, quadratic in perturbations: $\alpha$, $\beta$ and $\varepsilon$. At the quadratic level, \eqref{action} in fact captures all single-field models with scalar perturbations obeying the usual, phonon-like dispersion relations 
\beq
\label{phdr}
\omega = c_s k~,
\eeq
 at energy scales of order $H$, and our results apply to any theory with the latter property. 

Current data still allows for an appreciable range of parameters for the EFTI \eqref{action}, leaving room for detectable non-Gaussianity. Reducing the existing upper bound on the tensor-to-scalar ratio by less than an order of magnitude,
\beq
\label{smallr}
r< 10^{-2}~, 
\eeq
would however put the theories that predict the tensor and the scalar tilts of the same order ($ |n_T| \sim |n_s-1|\sim \epsilon$) in a serious tension with experiment. The model that has a slightly better chance of surviving the bound \eqref{smallr} is slow-roll inflation with weakly broken galileon symmetry (SRWBG) of Ref. \cite{Pirtskhalava:2015zwa}. Unlike the canonical slow-roll models with plateau-like potentials famously consistent with \eqref{smallr}, this model can be consistent with the latter constraint even if driven by a convex potential, as well as give rise to somewhat strongly coupled and highly non-Gaussian ($|f_{\rm NL}|\sim 1-20$) scalar perturbations \cite{Pirtskhalava:2015zwa}.

Set aside inflationary model-building, the results of this paper can be considered an extension of the calculations by Maldacena \cite{Maldacena:2002vr} and Chen et al. \cite{Chen:2006nt} of the full cubic action---including mixing with gravity---for the comoving curvature perturbation $\zeta$~. The first of these references has dealt with the case of canonical slow-roll inflation; Ref. \cite{Chen:2006nt}, on the other hand, has generalized the analysis to the case when arbitrary terms of the form $P\(\phi,(\p\phi)^2\)$ are present in the inflaton action. This is equivalent to considering operators of the form $(\delta N)^{n}$ in the EFT of inflationary perturbations. Here we further generalize the analysis of Chen et al. to the case when the operators of the form $(\delta N)^n \delta E$ become relevant in the unitary gauge EFT, which is naturally true for the broad class of theories with weakly broken galileon invariance. A calculation somewhat related to ours (although performed in a different language) has been carried out in  Ref. \cite{Kobayashi:2011pc}. That reference has however concentrated on the concrete model of \textit{G-inflation}, and the corresponding calculation is less general than the effective field theory treatment we adopt here. We find the EFT approach rather convenient since it captures an overwhelming majority of the existing single-field models, as well as it makes transparent the above-described interesting features  in Eqs. \ref{largeng} and \eqref{1overr}. Our presentation is close in spirit to those of \cite{Maldacena:2002vr,Chen:2006nt}, and in the appropriate limits our results are in agreement with the results of these references, as well as of Ref. \cite{Kobayashi:2011pc}.

The rest of the paper is organized as follows. In Section \ref{sec2} we study the dynamics of scalar perturbations in the theory specified by the action \eqref{action}, and identify the full set of regions in the parameter space characterized by significantly subluminal/non-Gaussian scalar perturbations. Furthermore, we categorize the list of existing, as well as novel, models of inflation with large non-Gaussianity according to which one of these regions they fall into. In Section \ref{sec3}, we explore various constraints---both theoretical and experimental---that these theories are subject to. Finally, in Section \ref{sec4}, we conclude. Various technical details, that would overwhelm the main body of the text, are collected in the three appendices. 

\section{Dynamics of scalar perturbations}
\label{sec2}

We start out with a brief account of the full analysis---including mixings with gravity---of scalar perturbations in the theory specified by the action \eqref{action}. To this end, we closely follow and generalize Maldacena's calculation \cite{Maldacena:2002vr} for canonical slow-roll inflation (see also Ref. \cite{Chen:2006nt}.)

The gauge freedom that remains after fixing to the unitary gauge, $\delta\phi(x,t)=0$, can be used \cite{Maldacena:2002vr} so as to put the three-dimensional metric into the following form, $g_{ij} = a^2 e^{2\zeta}( \delta_{ij}+h_{ij})$~. In this gauge, the scalar and tensor perturbations are captured by $\zeta$ and $h_{ij}$ respectively (the lapse and shift variables are non-dynamical, and can be expressed in terms of the rest of the degrees of freedom using their equations of motion.) We will be exclusively interested in the $n-$point functions of $\zeta$, so we ignore $h_{ij}$ altogether in the remainder of this work. 
Integrating out the (scalar parts of) perturbations $\delta N$ and $N_i\equiv \p_i\psi$ from the Hamiltonian and momentum constraint equations, one obtains the effective action for the only remaining scalar degree of freedom, $\zeta$~. The procedure is quite tedious, but straightforward, and is outlined in Appendix \ref{appa}; the final result for the quadratic action reads
\begin{equation}
\begin{split}
S^{(2)} 	= \int\D^4x \, a^3 \mathcal{N} \bigg[
	\dot{\zeta}^2 - c_s^2\frac{(\partial_i\zeta)^2}{a^2}\bigg]~ ,
\label{mg-a2}
\end{split}
\end{equation}
where the explicit expressions for the kinetic normalization factor  $\mathcal{N}$ and the speed of sound $c_s$ are given in Eqs. \eqref{mg-N} and \eqref{mg-cs} of Appendix \ref{appa}.
In terms of the dimensionless parameters of Eq. \eqref{ps}, these are
\beq
\begin{split}
\mathcal{N} &= \mpl^2 ~\frac{3\alpha(\alpha-2)+\beta+\varepsilon}{(\alpha-1)^2}~,\\
c_s^2 &= \frac{\alpha (1-\alpha)+\varepsilon}{3\alpha^2-6\alpha+\beta+\varepsilon}~,
\end{split}
\label{cssq}
\eeq
We have neglected the $\p_t\hat M_1^3$ term in the expression for $c_s^2$, Eq. \eqref{mg-cs}. 
More generally, we will neglect time derivatives of all the free EFT coefficients (in the second line) of the action \eqref{action} throughout this work\footnote{One can expect that the time dependence of any EFT coefficient $M^n$ in \eqref{action} satisfies $\p_t M^n\sim \varepsilon' H M^n\ll HM^n$ on a quasi-de Sitter space, where $\varepsilon'$ is slow-roll suppressed. As a result, the effects associated with non-zero time derivatives of these coefficients are generically suppressed with respect to their leading-order effects. We have explicitly checked this fact for the cases we consider below.}.
This assumption will not change our conclusions in any appreciable way, and we choose to make it in order to keep the presentation as simple as possible. 

In what follows we will be mostly exploring the properties of the theory in the $\alpha - \beta$ plane, fixing the rest of the parameters to some constant values. The reason is that it is precisely these two parameters---along with the slow-roll parameter $\varepsilon$---that determine the properties of the quadratic perturbation Lagrangian, see Eqs. \eqref{mg-a2} and \eqref{cssq}. As a result, two out of the three constraints that we'll impose below, namely those stemming from stability/subluminality and the tensor-to-scalar ratio, are unambiguously determined by $\alpha,~\beta$ and $\varepsilon$~. As to the third constraint arising due to non-Gaussianity, the latter three parameters do also contribute to $f_{\rm NL}$~. Whenever this contribution becomes too large, one can turn on non-zero values for $\gamma$ and $\delta$ to bring $f_{\rm NL}$ back within the observational limits. However, barring such an adjustment of parameters, it is a rather interesting fact that the set of just three numbers, $\alpha,~\beta$ and $\varepsilon$, can tell us a great deal about the theoretical and experimental status of an overwhelming majority of single-field inflationary theories. 

In fact, the latter set of parameters determines the phenomenology of any single-field model of inflation in which the scalar perturbations are characterized by the usual, phonon-like, dispersion relation \eqref{phdr}
at freezout of the CMB modes. The reason is as follows. At the quadratic order in perturbations, there are only two additional operators one can add to the Lagrangian \eqref{action} to be consistent with Eq. \eqref{phdr}. These are $\sqrt{-g}~\(\delta E^i_j\delta E^j_i-(\delta E)^2\)$ and $\sqrt{-g} ~\delta N \upleft{3}{R}$~. Both of these operators, however, are redundant and can be removed by a perturbative field redefinition \cite{Creminelli:2014wna}, so at least the quadratic piece of our action \eqref{action} is very generic. Since, as remarked above, the majority of the constraints on the EFT \eqref{action} stem precisely from the \textit{quadratic} Lagrangian, our analysis should (at least qualitatively) capture phenomenology of any model satisfying \eqref{phdr}. 

Particularly interesting regions in the parameter space are the ones that correspond to the (squared) speed of sound, $c_s^2$~, becoming small. The reason is that this generically enhances the couplings of scalar perturbations, giving rise to fairly large values of non-Gaussianity, potentially relevant for observations.
One can see, from Eq. \eqref{cssq}, that there are a number of ways to make the sound speed suppressed. We will divide these into the following broad classes, which encompass most of the single-field models capable of generating large non-Gaussianity.

\subsection{DBI, and related models}
\label{dbipar}
 
In the unitary gauge, these models give rise to the following relations between the dimensionless parameters \cite{Alishahiha:2004eh, Chen:2006nt}
\beq
\big \{ \alpha, ~\gamma \big \} \sim 0~, \qquad  \big\{ ~|\beta|, ~|\delta|~ \big \}\gsim \varepsilon~,
\eeq
implying the following behaviour of the amplitude of non-Gaussianity for strongly subluminal perturbations \cite{Chen:2006nt}
\beq
\label{fnldbi}
f_{\rm NL}\propto \frac{1}{c_s^2}~.
\eeq
The two non-zero coefficients $\beta$ and $\delta$ in fact parametrically exceed the slow-roll parameter $\varepsilon$ in the small $c_s^2$ limit:
$\beta \propto \varepsilon/c_s^2$, and $\delta \propto \varepsilon/c_s^4~$.
However, current experimental bounds imply $c_s^2\gsim \varepsilon$ for the DBI model \cite{Ade:2015lrj}, so perhaps the optimal values for these parameters to keep in mind are
\beq
|\beta|\lsim 1, \qquad |\delta|\sim \frac{\beta}{c_s^2}~.
\eeq

\subsection{G-inflation/Galileon inflation}
\label{ginfpar}
 
In this category, we collect inflationary theories characterized by the dimensionless parameters of \eqref{action} satisfying the following conditions
\beq
\label{ginfpar1}
\big \{~ |\beta| ~, |\gamma|, ~|\delta|~\big \} \sim 1~, \qquad |\alpha |< 1 ~,
\eeq
so that the subluminal limit corresponds to
\beq
\label{ginf1}
c_s^2\sim \frac{\alpha}{\beta} < 1~.
\eeq 
This can be the case in G-inflation \cite{Kobayashi:2010cm}, Galileon inflation \cite{Burrage:2010cu} and, more generally, in the kinetically driven phase of theories with weakly broken galileon symmetry \cite{Pirtskhalava:2015nla}.
For a suppressed speed of sound, the amplitude of non-Gaussianity grows similarly to DBI models
\beq
f_{\rm NL}\propto \frac{1}{c_s^2}~.
\label{fnldbi1}
\eeq
It has been noticed by Burrage et al. \cite{Burrage:2010cu}, however, that in the unitary gauge, the most general theory of Galileon inflation introduces extra cubic operators on top of the ones present in \eqref{action}, such as $\sqrt{-g}~\delta N\(\delta E^i_j\delta E^j_i-(\delta E)^2\)$ for example (with an order-one coefficient in Planck units), and these can result in a faster growth 
\beq
f_{\rm NL}\propto \frac{1}{c_s^4}~.
\label{nggi}
\eeq
We have not included such operators in our analysis. The major reason is that, as we will see shortly, constraints on various inflationary models described by the EFT \eqref{action} are already quite strong regardless of the cubic operators; including the latter can loosen the constraints due to non-Gaussianity\footnote{Being manifestly at least cubic in perturbations, these operators only affect the bispectrum from the set of physical quantities we consider below.}, but only at an expense of tuning/cancellations. 
As to the rest of the constraints of the next section, they do apply equally well to the most general theory of G-/Galileon inflation, as discussed above Eq. \eqref{phdr}. 

\subsection{Kinetically driven inflation with WBG symmetry }
\label{kwbgpar}

This model is characterized by
\beq
\label{kdwbg}
\alpha \simeq 1~, \qquad \big\{~ |\beta|,~|\gamma|,~|\delta| ~\big \} \sim 1~,
\eeq
and arises in the context of general theories with weakly broken galileon symmetry \cite{Pirtskhalava:2015nla}, of which the \textit{Covariant Galileon} \cite{Deffayet:2010qz} behind Galileon inflation is a particular case. 
The values of the parameters in \eqref{kdwbg} can in fact also arise in Galileon inflation \cite{Burrage:2010cu}; unfortunately, however, the corresponding regime can not be captured by the decoupling limit analysis adopted in the latter reference. We have chosen to present this theory as a separate class, since it gives rise to the fastest growth of non-Gaussianity with the small speed of sound,
\beq
\label{ngfast}
f_{\rm NL}\propto \frac{1}{c_s^6}~,
\eeq
well within the regime of validity of the low energy EFT. We give a detailed account of the kinetically driven inflation with weakly broken galileon symmetry (KWBG) in Appendix \ref{appc}. The expression in Eq. \eqref{ngfast} arises as a result of subtle effects associated with mixing of the adiabatic perturbations with the metric degrees of freedom, partially explaining why it has gone unnoticed in the literature. 

\subsection{Slow-roll inflation with WBG symmetry}
\label{srwbgpar}

This class of theories was introduced in Ref. \cite{Pirtskhalava:2015zwa} and gives rise to the values of the dimensionless parameters of the EFT  \eqref{action} suppressed by slow-roll 
\beq
\label{srwbg}
\big \{~|\alpha|,~ |\beta|,~|\gamma|,~|\delta|~\big \} \sim \varepsilon ~.
\eeq
The speed of sound of scalar perturbations is generically order-one (but not necessarily strictly one, in contrast to canonical slow-roll inflation.) The background evolution, along with all background characteristics (spectral tilt, number of e-folds from freezeout of the CMB modes until the end of inflation, spectrum of gravitational waves, etc.) are all parametrically similar to garden-variety slow-roll models. What's different, though, is that the scalar perturbations can be more strongly coupled than in canonical slow-roll inflation -- in a way that, nevertheless, allows to keep control over the derivative expansion. This leads to an amplitude of non-Gaussianity that grows like
\beq
f_{\rm NL} \propto \frac{1}{c_s^4}
\label{fnlsrwbg}
\eeq
in the subluminal limit. Note that, while this behaviour is similar to \eqref{nggi}, the underlying models are very different: the SRWBG model is a minimal deformation of slow-roll inflation, unlike Galileon inflation which describes a \textit{kinetically-driven} background. Moreover, as already mentioned above, the version of Galileon inflation described by the EFT \eqref{action} in fact yields a DBI-like growth, Eq. \eqref{fnldbi}, while the slow-roll theories with WBG invariance lead to \eqref{fnlsrwbg} already within the realm of the EFT \eqref{action} -- i.e. without the need of having extra cubic operators. The strong dependence of non-Gaussianity on the speed of sound allows to generate appreciable values for $f_{\rm NL}$ even for mildly subluminal perturbations, arising for
\beq
\label{srwbgngcond}
\alpha \simeq - \varepsilon 
\eeq
in the SRWBG model (see Eq. \eqref{cssq}.) 

Having classified various alternatives to canonical slow-roll inflation according to the way they generate large non-Gaussianity, we proceed to explore the theoretical and experimental constraints on the general EFT parameter space \eqref{ps} in the next section.  

\section{Constraints}
\label{sec3}

There is a number of constraints---both theoretical and experimental---that the models \ref{dbipar} through \ref{srwbgpar} discussed in the previous section are subject to. 
Above all, there is a constraint expressing the absence of negative norm states (or, alternatively, boundedness from below of the Hamiltonian) and of gradient instability. Theories, that do not satisfy this constraint can hardly be made sense of. A somewhat less sharp constraint comes from demanding the absence of superluminal scalar perturbations. While it is not fully clear whether superluminality within a low-energy EFT is unconditionally unacceptable, there are good reasons to believe that at the very least it is inconsistent with the standard properties (Lorentz-invariance, locality, analyticity, etc.) of a hypothetical UV completion \cite{Adams:2006sv}. To be on the safe side, we will thus demand that the scalar excitations are subluminal as well. 
The above considerations then summarize into the following conditions on the dimensionless parameters of the theory
\beq
\mathcal{N}>0~, \qquad 0<c_s^2\leq 1~.
\eeq

Furthermore, an important role for our analysis will be played by the current limits on the amplitude of primordial gravitational waves.  The scalar-to-tensor ratio, $r$, can be readily read off the quadratic $\zeta$ action, Eq. \eqref{mg-a2},
\beq
\label{stratio}
r = 16~ \frac{\mathcal{N} c_s^3}{\mpl^2}=16~ \frac{\varepsilon+\alpha-\alpha^2}{(\alpha-1)^2}~\sqrt{\frac{\varepsilon+\alpha-\alpha^2}{3\alpha^2-6\alpha+\beta+\varepsilon}}~.
\eeq
In the slow-roll limit, $\alpha=\beta=0$, this reduces to the familiar expression $r_{\rm sr}=16 \varepsilon$, while for DBI inflation ($\alpha=0$), Eqs. \eqref{stratio} and \eqref{cssq} yield, $r_{\rm DBI} = 16\varepsilon c_s$~.

Last but not least, we will impose the experimental limits on primordial non-Gaussianity. The full calculation of the scalar bispectrum for the theory \eqref{action} is presented in Appendices \eqref{appa} and \eqref{appb}, and we will not reproduce it here. We define the (shape-independent) measure for the strength of non-Gaussian effects, $f_{\rm NL}$, in the following way
\beq
\label{fnl1}
f^{\rm}_{\rm NL} = \frac{5}{18} \frac{B_\zeta (k,k,k)}{P_\zeta(k)^2}~,
\eeq
where the bispectrum $B_\zeta$ can be found from the three-point function of $\zeta$
\beq
\langle \zeta(\vec k_1)\zeta(\vec k_2)\zeta(\vec k_3)\rangle=(2\pi)^3\delta^{(3)}\(\sum_i \vec{k}_i\)B_\zeta(k_1,k_2,k_3)~.
\eeq
As to the shape of non-Gaussianity, it is generically well-approximated by the equilateral one \cite{Creminelli:2005hu} in the theory \eqref{action}.

The precise expression for $f_{\rm NL}$ in terms of the dimensionless parameters of the action \eqref{action} (including the slow-roll parameter $\varepsilon$) is not particularly illuminating. As discussed in the previous section, in different regions of the parameter space corresponding to significantly subluminal scalar perturbations, $f_{\rm NL}$ acquires simple leading behaviour of the type $f_{\rm NL}\propto 1/c_s^{2p}$, with $p=1,2$ or $3$. Moreover, for non-zero $\gamma$ and $\delta$, there is a "$f_{\rm NL}\propto 1/r$ effect", mentioned below Eq. \eqref{1overr}.

\begin{figure}
\centering
\includegraphics[width=.45\textwidth]{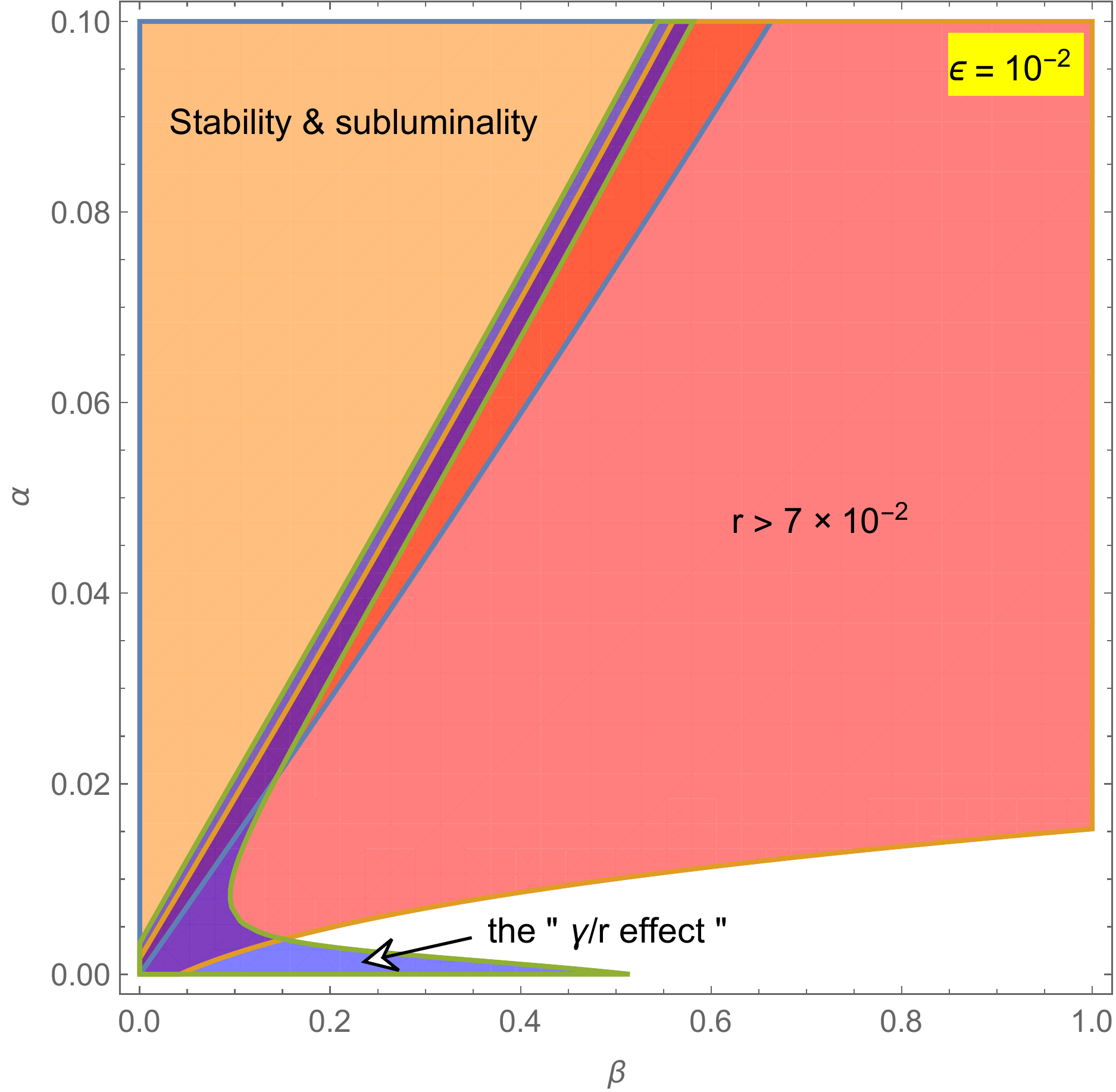} \quad
\includegraphics[width=.45\textwidth]{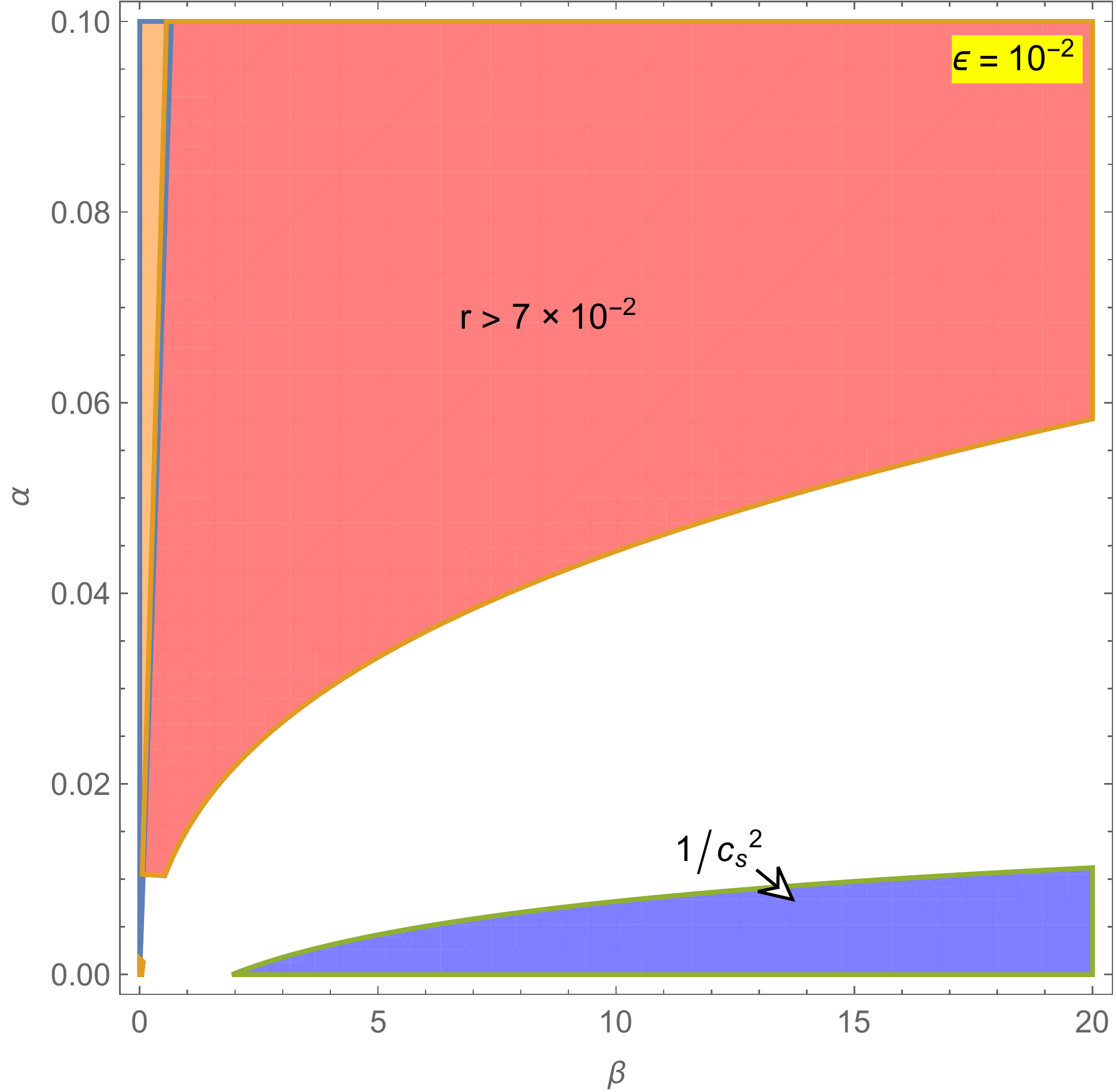} \\
\includegraphics[width=.45\textwidth]{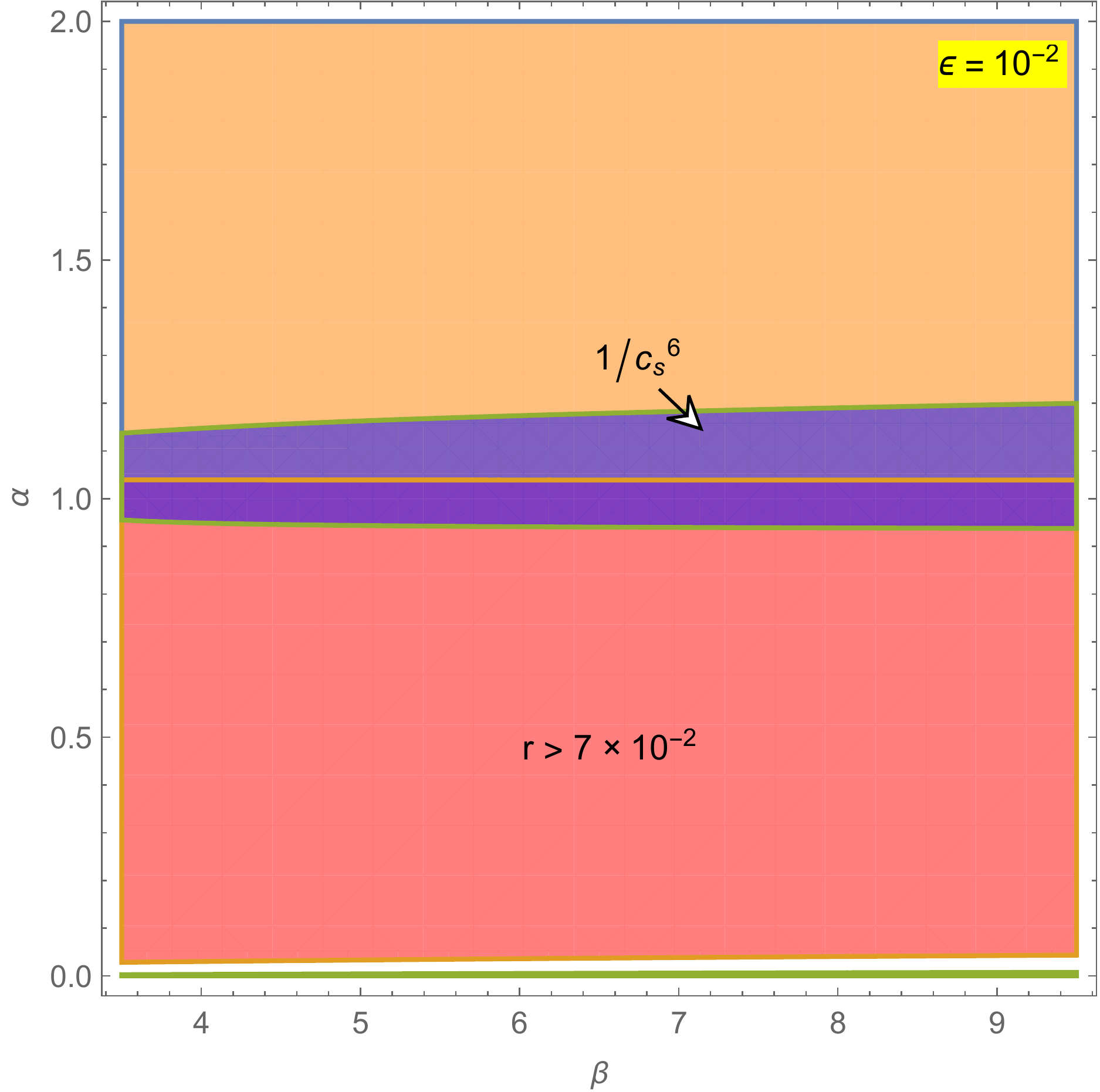} \quad
\includegraphics[width=.45\textwidth]{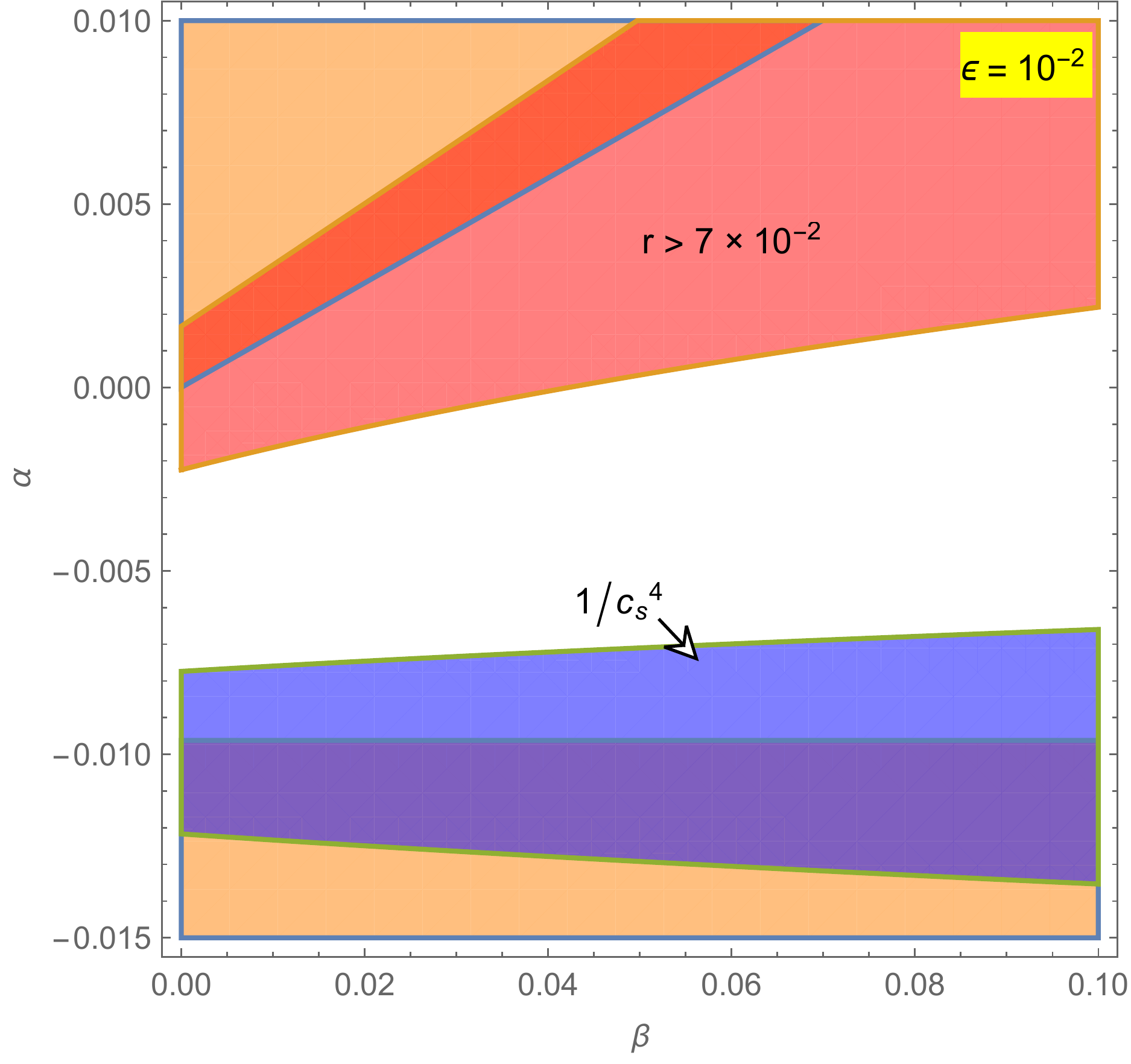}
\caption{Various regions of the parameter space, excluded by the requirements of stability \& subluminality (orange), current limits on the amplitude of primordial gravitational waves (red), and on non-Gaussianity (blue). The slow-roll parameter has been fixed to $\varepsilon=10^{-2}$, and the red and blue bands correspond to regions excluded respectively by the bounds $r<0.07$ and $-50<f_{\rm NL}<50$. The parameters $\gamma$ and $\delta$ have been chosen to vanish everywhere except the upper left panel, where they have been set to $\gamma=\delta=5$. }
\label{fig:1}
\end{figure}
\begin{figure}
\centering
\includegraphics[width=.45\textwidth]{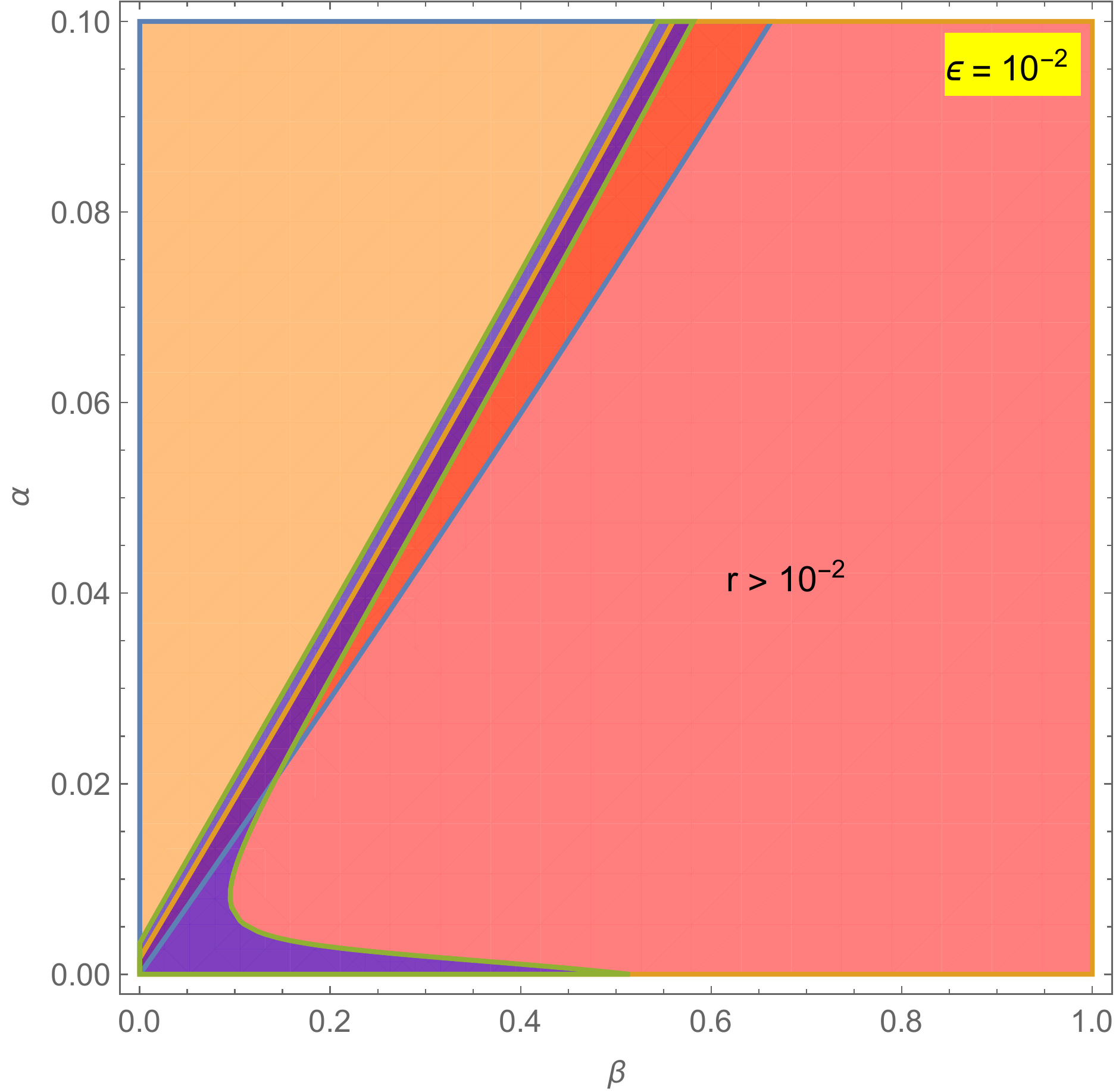} \quad
\includegraphics[width=.45\textwidth]{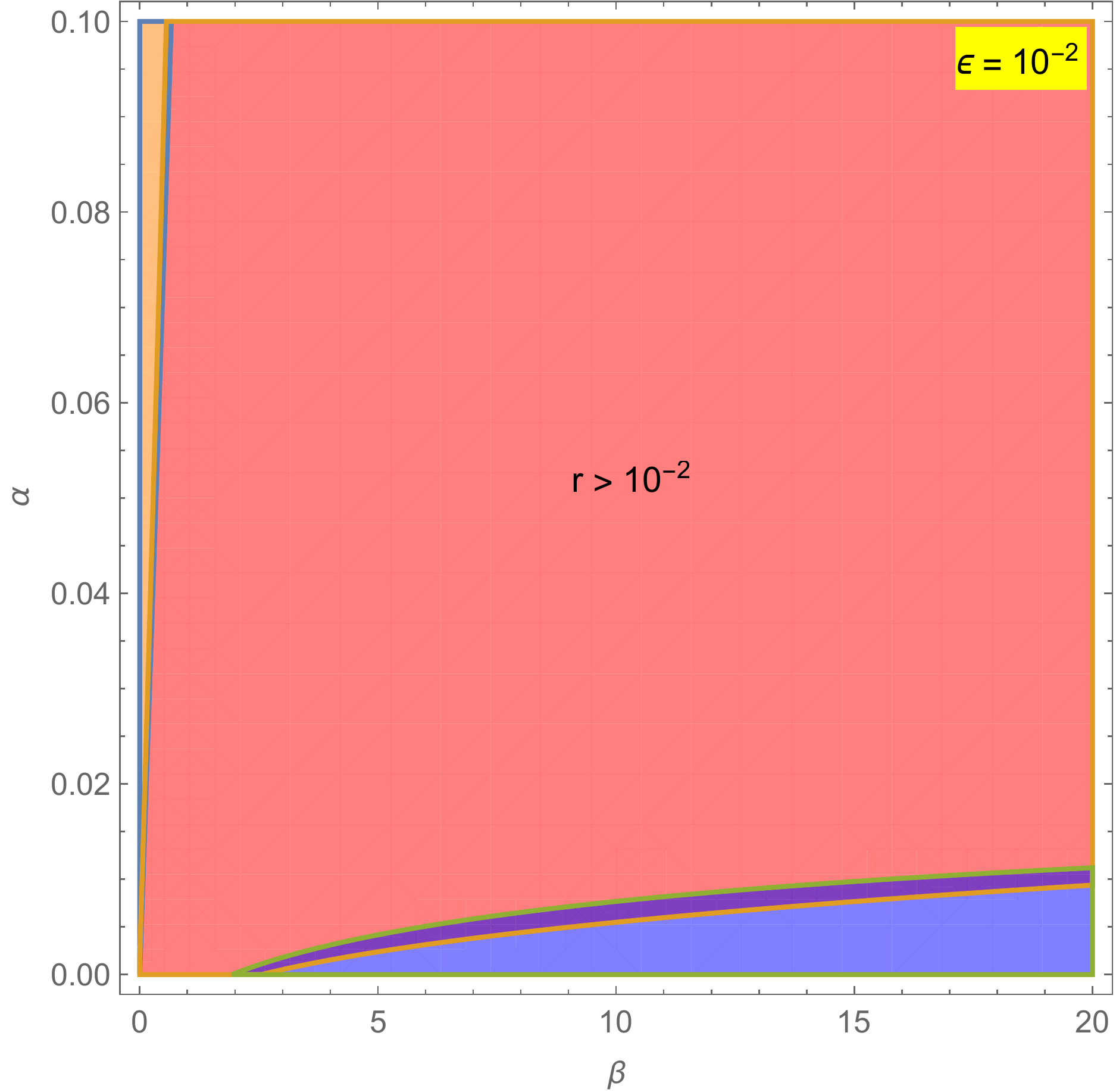} \\
\includegraphics[width=.45\textwidth]{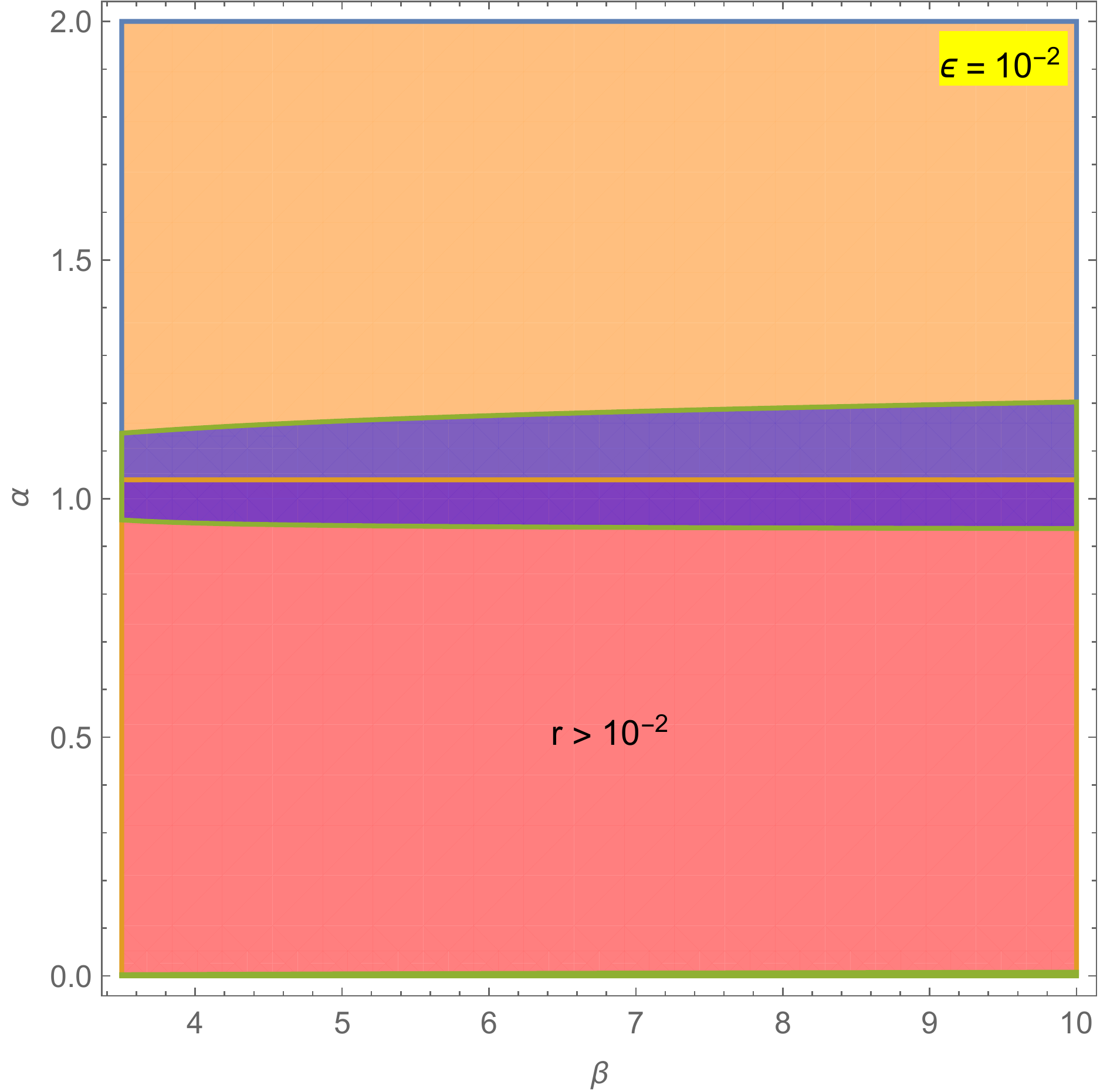} \quad
\includegraphics[width=.45\textwidth]{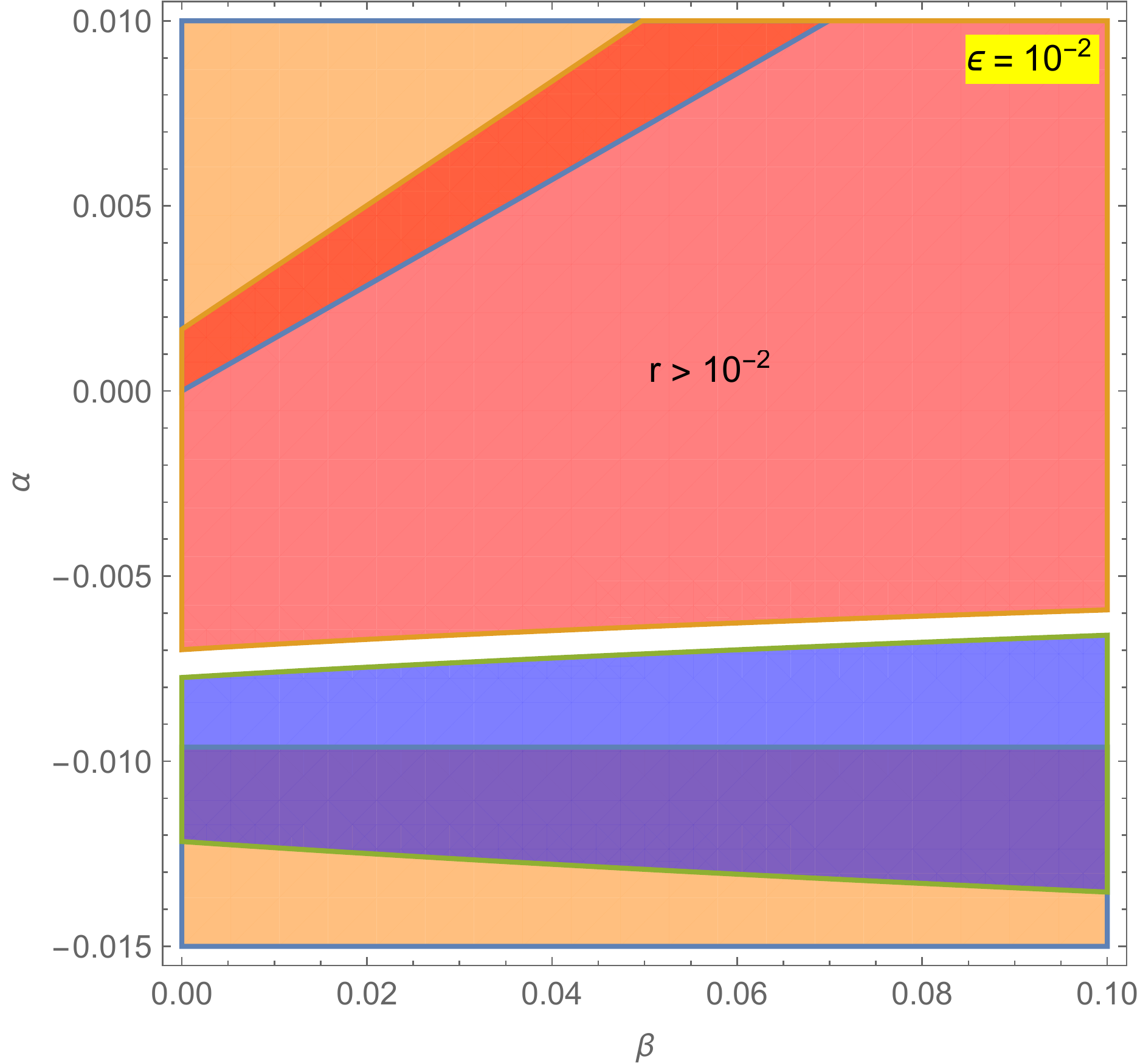}
\caption{Here we illustrate the exact same situation as in Fig. \ref{fig:1}, with the only difference in the exclusion from the tensor-to-scalar ratio: the red band now corresponds to the regions, where $r>10^{-2}$. }
\label{fig:2}
\end{figure}

We will ultimately project the parameter space onto the $\alpha - \beta$ plane, so some input regarding the magnitude of the slow-roll parameter $\varepsilon$ is needed. The measured tilt of the primordial scalar spectrum $n_s$ suggests that $\varepsilon\lsim |n_s-1|\sim 10^{-2}$. The latter bound is saturated for many inflationary models -- e.g. those with convex potentials. On the other hand, there are models characterized by plateau-like potentials such as Starobinsky's $R^2$ inflation \cite{Starobinsky:1980te}, or the so-called IR DBI inflation \cite{Chen:2005fe}, where $\varepsilon$ can be much smaller than the scalar tilt. In order to capture both classes of models, we will assume two values for the slow roll parameter in our analysis: $\epsilon = 10^{-2}$ and $\epsilon\sim 0$ (the latter precisely defined below.)  

The three constraints discussed above lead to an interesting interplay, in many cases excluding complementary regions of the parameter space. Consider, for example, a DBI-like model with a generic power-law potential, so that $\varepsilon \sim 10^{-2}$~.
 The tensor-to-scalar ratio,  $r_{\rm DBI} = 16\varepsilon c_s$, has to be below $\sim 0.1$ according to the current experimental limits \cite{Ade:2015lrj}, requiring a somewhat suppressed speed of sound. On the other hand, significantly suppressing $c_s$~, one runs into tension with the current limits on non-Gaussianity, in accord with Eq. \eqref{fnldbi}. The precise constraint for DBI models is $c_s\gsim 0.1$ \cite{Ade:2015ava}. This means that measuring $r\sim10^{-2}$ would rule out the given class of theories. In contrast, DBI theories driven by plateau-like potentials like the IR model of Ref. \cite{Chen:2005fe}, are characterized by $\varepsilon \ll 10^{-2}$ and therefore have a better chance of being consistent with a small tensor-to-scalar ratio. 
 
\begin{figure}
\centering
\includegraphics[width=.45\textwidth]{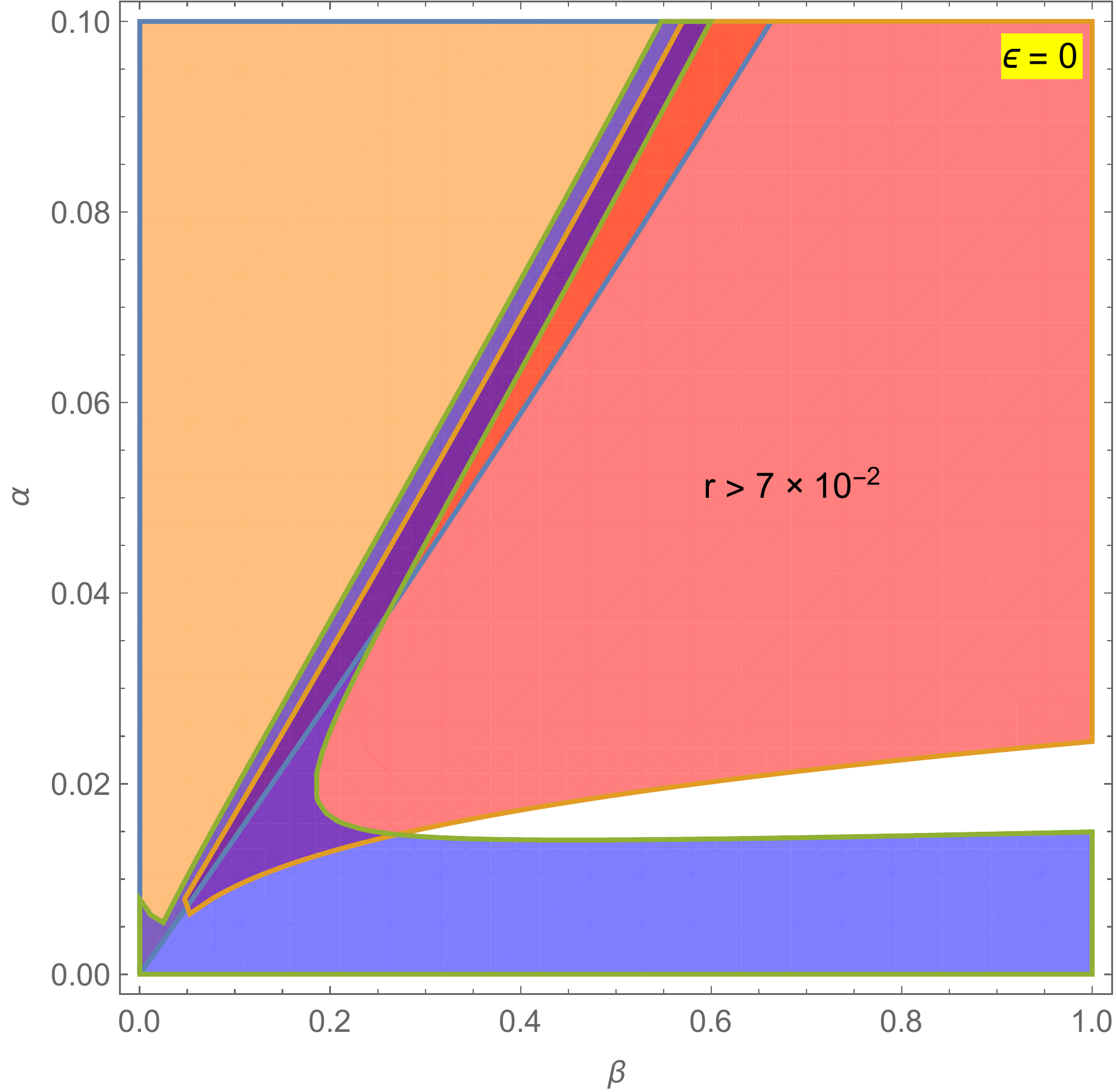} \quad
\includegraphics[width=.45\textwidth]{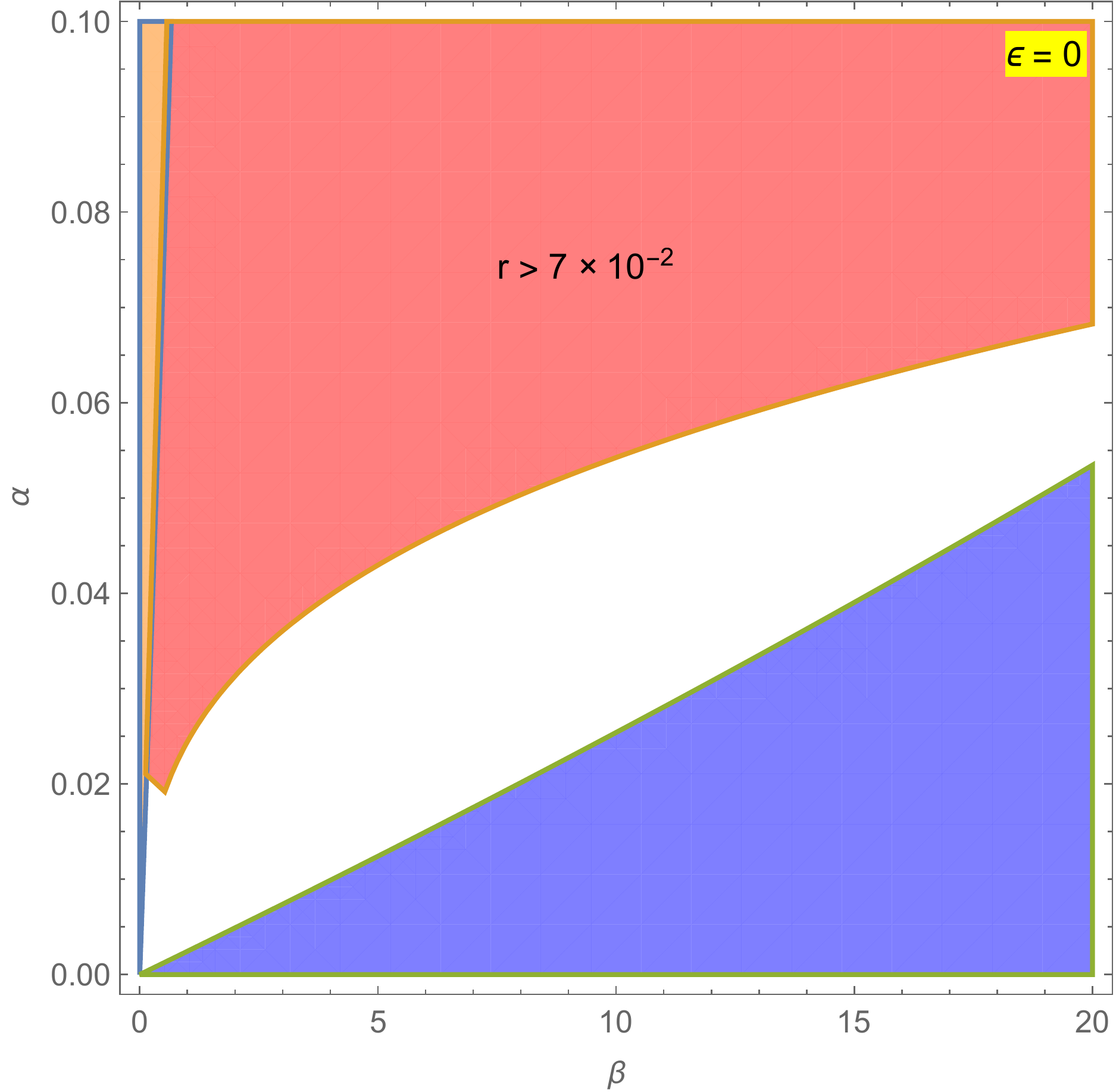} \\
\includegraphics[width=.45\textwidth]{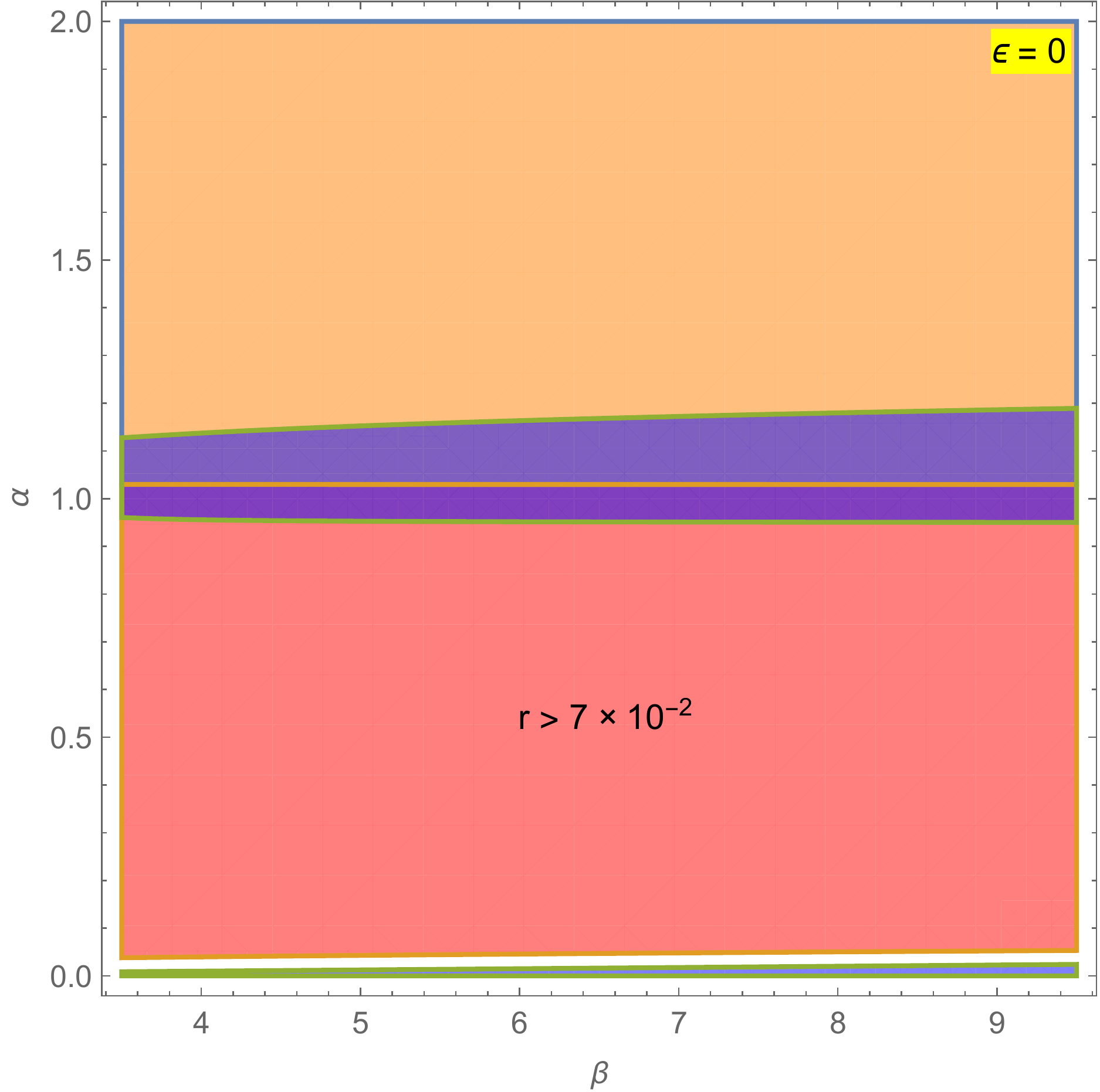} \quad
\includegraphics[width=.45\textwidth]{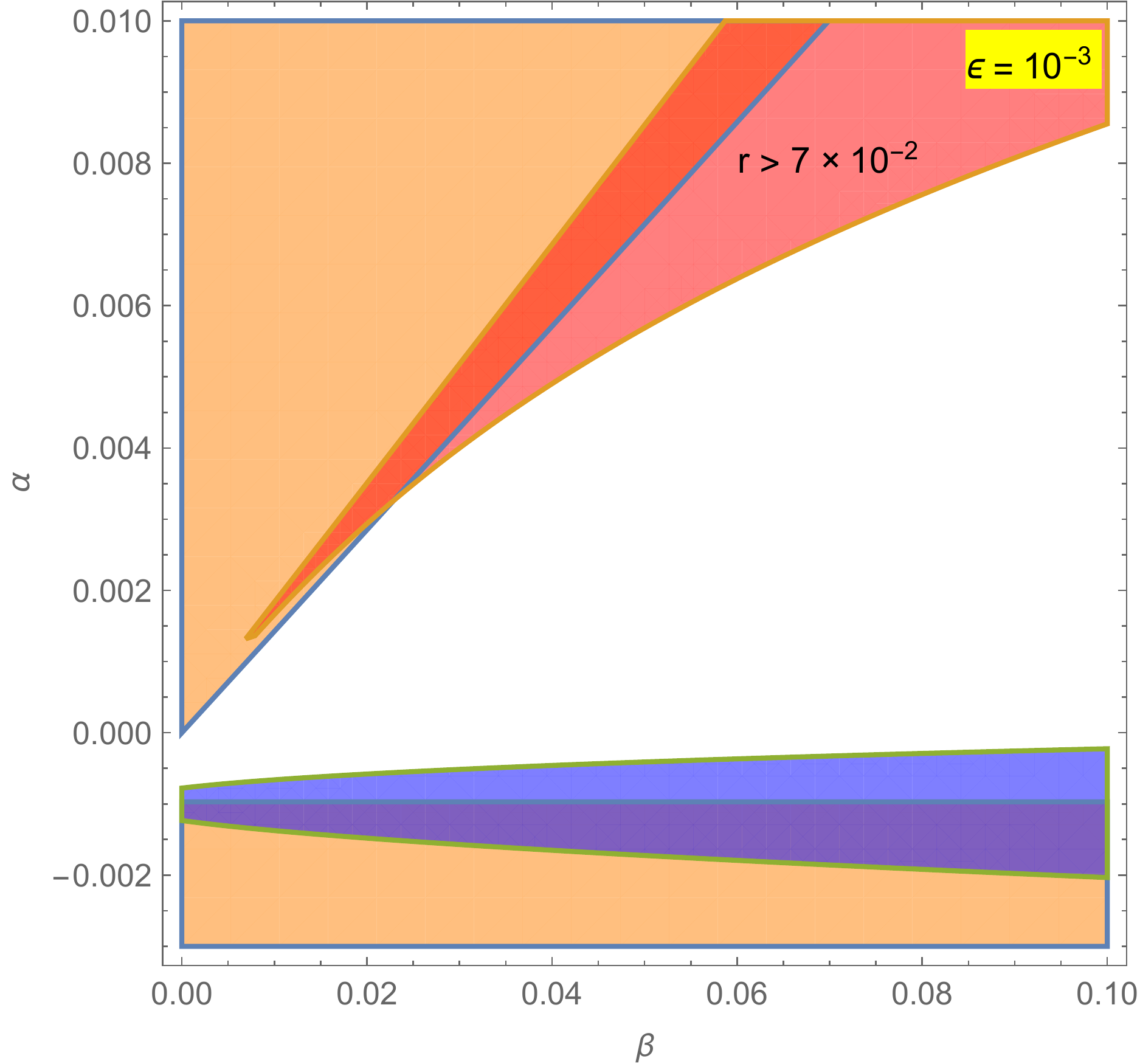}
\caption{Various regions of the parameter space corresponding to models discussed in Sec. \eqref{sec2}. The red band shows the regions excluded by requiring $r<7\times 10^{-2}$, while $\epsilon = 0$ in all panels but the lower right one, which has $\epsilon = 10^{-3}$. }
\label{fig:3}
\end{figure}

\begin{figure}
\centering
\includegraphics[width=.45\textwidth]{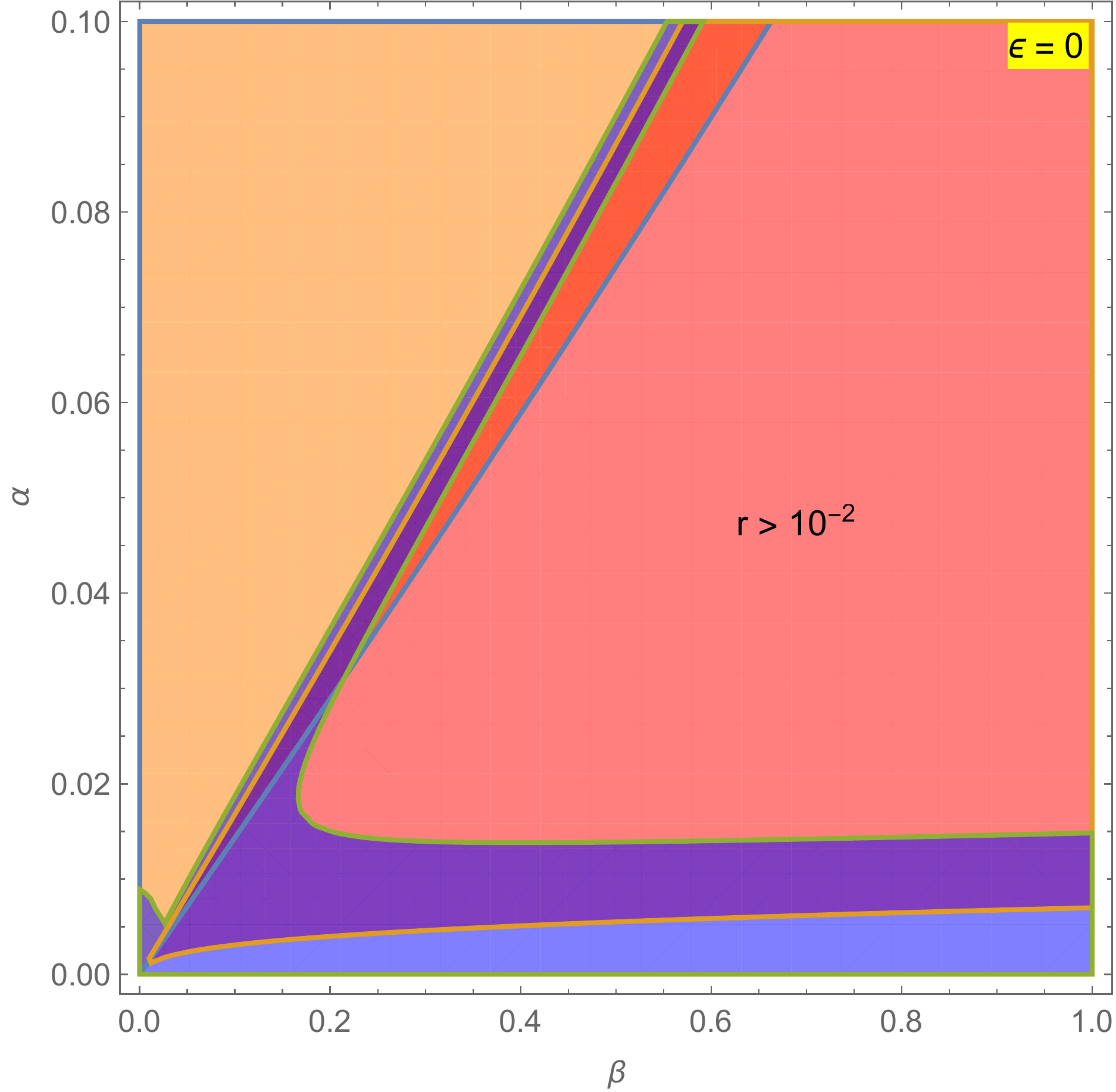} \quad
\includegraphics[width=.45\textwidth]{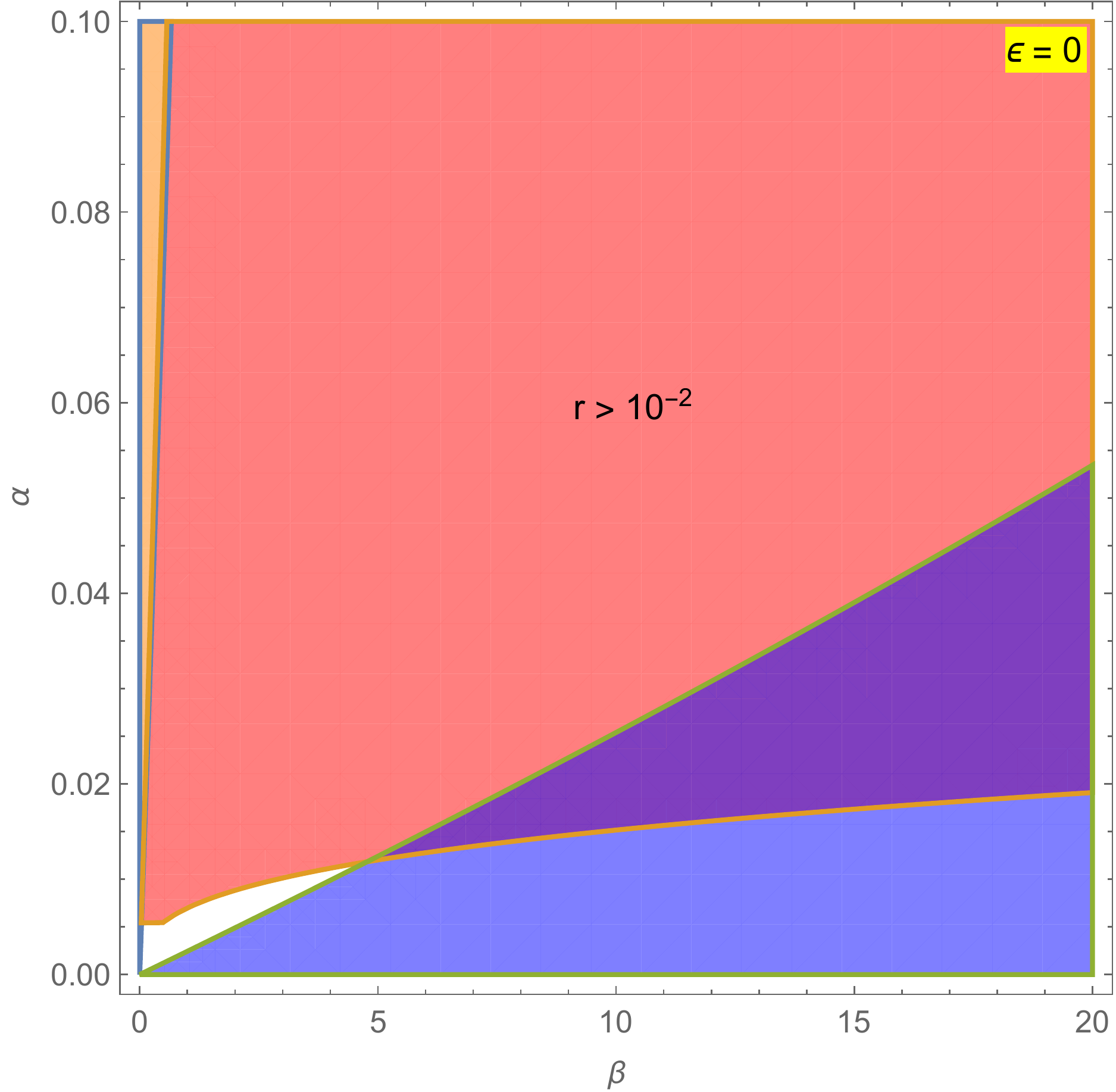} \\
\includegraphics[width=.45\textwidth]{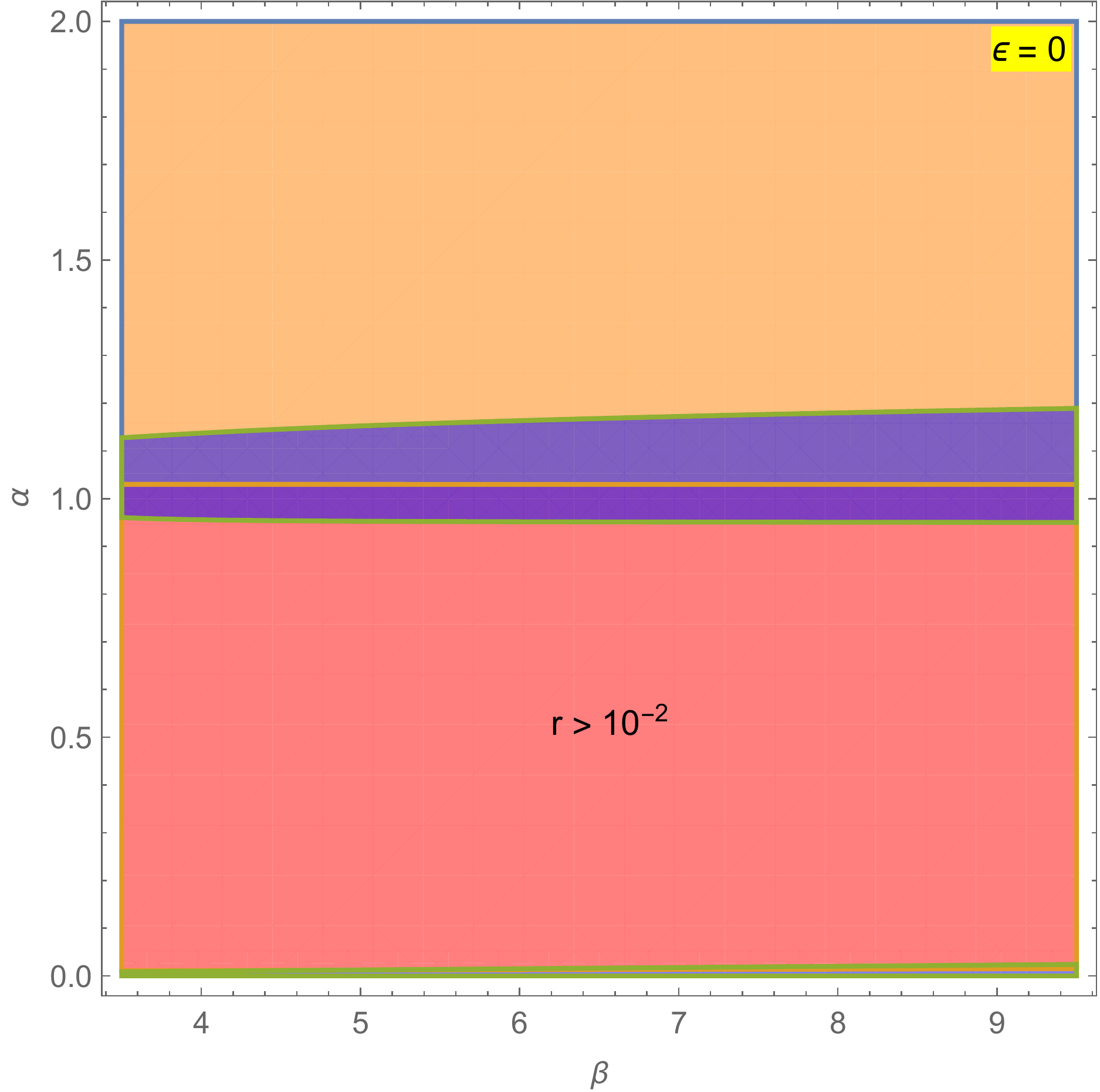} \quad
\includegraphics[width=.45\textwidth]{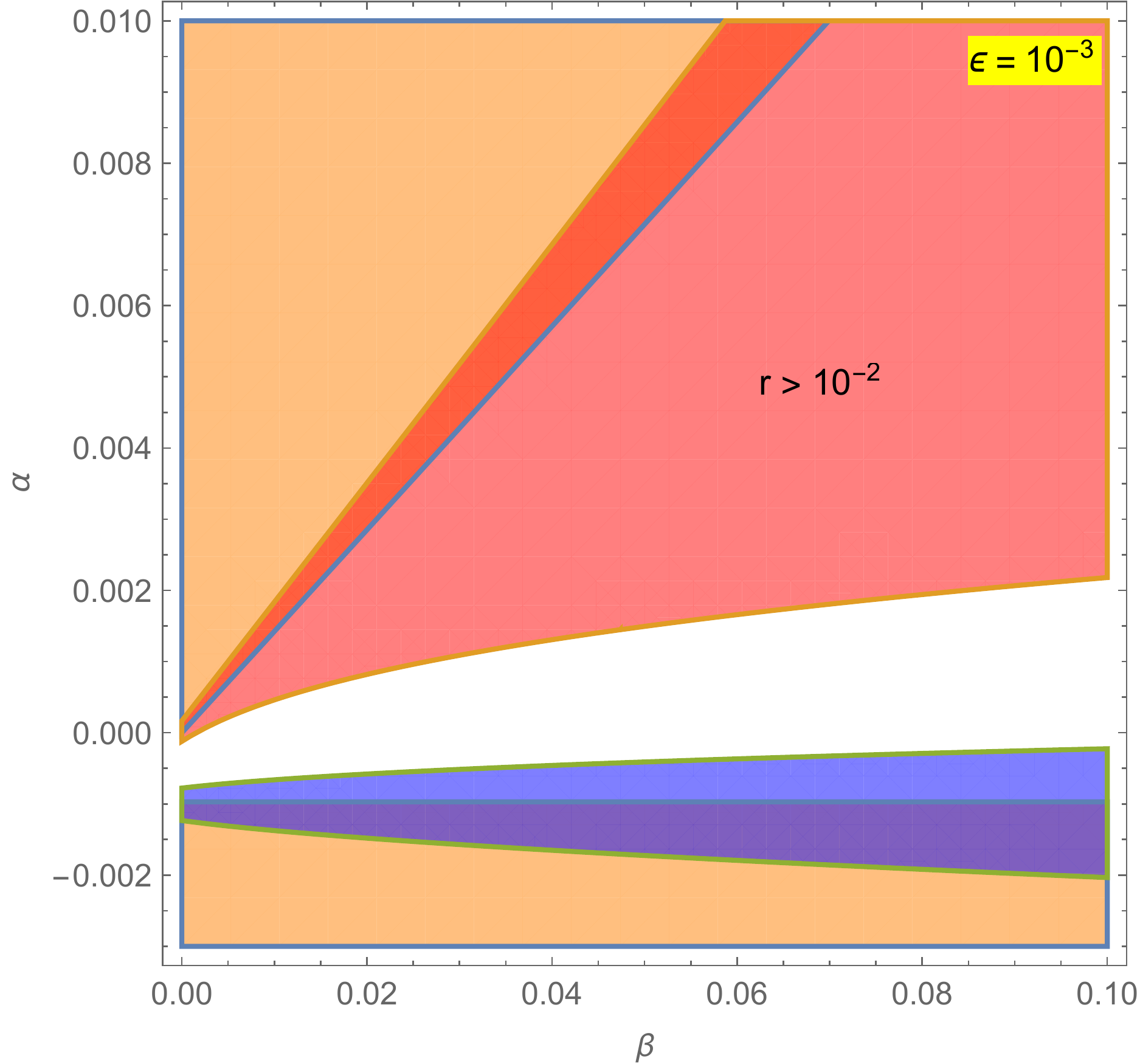}
\caption{Here we illustrate the exact same situation as in Fig. \ref{fig:3}, with the only difference in the exclusion from the tensor-to-scalar ratio: the red band now corresponds to the regions, where $r>10^{-2}$. }
\label{fig:4}
\end{figure}

The examples of exclusion plots for the general parameter space of interest are shown on Figs. \ref{fig:1} through \ref{fig:4}. On the first two of these figures, we assume $\varepsilon = 10^{-2}$, while the last two correspond to $\epsilon\sim 0$. Moreover, we require $f_{\rm NL}$ to be in the range $-50<f_{\rm NL}<50$, motivated by the current limits on equilateral non-Gaussianity \cite{Ade:2015ava}. The orange, red and blue regions depict parts of the parameter space excluded by instabilities and/or superluminality, limits on the tensor-to-scalar ratio, and on non-Gaussianity respectively. 

\subsection{Models with $\varepsilon \sim |n_s-1|$}

On Fig. \ref{fig:1} the red regions are the exclusion bands due to the present $95 \%$ C.L. bound on the amplitude of the primordial gravitational waves, $r<0.07$ \cite{Bicepnew}. One can see, that the data prefers significantly suppressed $\alpha$, effectively ruling out inflationary theories with $\alpha\gg 10^{-2}$ (this is a rather general result, true for all cases that we consider below \footnote{The allowed region can in fact reach out to $\alpha\simeq 0.1$, but this only happens for $\beta \sim 50$, casting shadow on the quantum stability of the corresponding theories.}.)

For $\alpha$ much smaller than $\varepsilon$, the boundary between the DBI and G-/Galileon inflation is blurred. An important discriminant that remains, though, is the fact that $\gamma$ can be much larger in the latter class of models. For this reason, we have chosen $\gamma=\delta=5$ in the upper left panel, which results in an additional exclusion region in G-/Galileon inflation due to large non-Gaussianity stemming from the "$1/r$" effect of Eq. \eqref{1overr} (from the two terms in this equation, only the one proportional to $\gamma$ contributes significantly.) Had we chosen $|\gamma| \ll1$ as in DBI inflation, this band would have completely disappeared from the plot. 
The regions of the $\alpha$--$\beta$ plane explored in the rest of the panels on Fig. \ref{fig:1} are not affected by $\gamma$ and $\delta$ for reasonable values of these parameters\footnote{For the upper right and the lower left panels, even setting $\gamma\sim\delta\sim 10$ has little effect on the exclusion regions. The lower right panel corresponds to the model \ref{srwbgpar} (SRWGB), where both of these parameters are naturally of order $\varepsilon$ and lead to negligible effects.} so we have set them to zero everywhere except the upper left one. 

The upper right panel of Fig. \ref{fig:1} shows the blue exclusion region due to non-Gaussianity in the $f_{\rm NL}\propto 1/c_s^2$ regime, characteristic of the models \ref{dbipar} and \eqref{ginfpar}. The blue band shown here appears for larger values of $\beta$, where the speed of sound becomes small enough (see Eq. \eqref{cssq})
\beq
c_s^2 \sim \frac{\mathcal{O}\(\alpha, \varepsilon\)}{\beta}~,
\eeq
so as to trigger the growth of $f_{\rm NL}$ according to Eqs. \eqref{fnldbi} and \eqref{fnldbi1}. 

Another part of the parameter space, corresponding to the model \ref{kwbgpar} (KWBG) is shown on the lower left panel. One can see, that the region excluded by non-Gaussianity due to the abrupt growth $f_{\rm NL}\propto 1/c_s^6$ is concentrated around $\alpha\simeq 1$ (as expected from Eq. \eqref{kdwbg}.) Unfortunately, this model is already ruled out by the limits on the primordial gravitational waves, combined with the theoretical requirements of stability and subluminality.

Finally, on the lower right panel, we zoom onto the parameter space corresponding to the slow-roll WBG model of Ref. \cite{Pirtskhalava:2015zwa}, discussed in Sec. \ref{srwbgpar}. One can see the exclusion band from non-Gaussianity around $\alpha\simeq -\epsilon$, corresponding to $f_{\rm NL}$ growing like $\sim 1/c_s^4$~. Just like in DBI/G-/Galileon inflation, there remains an appreciable portion of the parameter space still allowed by our constraints with $r\lsim 0.07$, including regions of the $\alpha$--$\beta$ plane characterized by detectable non-Gaussianity. We stress again, however, that the SRWBG model, being in  a well-defined sense a minimal deformation of canonical slow-roll inflation, is very different from the rest of the models considered in Sec. \ref{sec2}. 

While the current data still leaves some room for most of the models with $\varepsilon\sim |n_s-1|$, the situation can change dramatically if the upper limit on $r$ decreases to $r\lsim10^{-2}$ (which is less than an order of magnitude improvement in current precision.) The plots, corresponding to this case are shown on Fig. \eqref{fig:2}; one can see that the regions that were previously allowed are now fully covered by the exclusion bands from gravitational waves. The only region that still remains is a narrow band in the slow-roll model with weakly broken galileon invariance, shown on the lower right panel of Fig. \ref{fig:2}. 

The canonical models of slow-roll inflation sit at the origin of the $\alpha$--$\beta$ plane, and are of course not visible on our plots. Measuring $r\lsim 10^{-2}$ would rule out most of these, with an exception of models with plateau-like potentials, such as Starobinsky's $R^2$ inflation \cite{Starobinsky:1980te}. In these models, the tilt of the scalar spectrum is mostly determined by the second slow-roll parameter, $\eta_V\equiv \mpl^2 V''/V$, so that $\varepsilon$ can be much smaller than $|n_s-1|$ to suppress the tensor-to-scalar ratio. Needless to say, falling into the category of canonical slow-roll theories, $R^2$ inflation predicts undetectable non-Gaussianity, $f_{\rm NL}\sim 10^{-2}$ \cite{Maldacena:2002vr}. In contrast, (the non-canonical) slow roll inflation with WBG symmetry, \textit{even if driven by the simplest convex potentials} (with $\varepsilon\sim |n_s-1|$), does possess a parameter space consistent with tensor-to-scalar ratios as small as $r \lsim 10^{-2}$, as seen from the lower right panel of Fig. \ref{fig:2}. Moreover, close to the blue non-Gaussianity exclusion band, this model can generate detectable (equilateral) non-Gaussianity, $|f_{\rm NL}|\lsim 50$ \cite{Pirtskhalava:2015zwa}. 


\subsection{Models with $\varepsilon \ll |n_s-1|$}

It is well-known that the tensor-to-scalar ratio can be significantly suppressed (and therefore the bounds coming from this observable ameliorated) in models where the variation of the inflationary Hubble rate does not significantly contribute to the scalar tilt. A famous example are theories (slow-roll or not) driven by plateau-like potentials. To capture this class of models, we repeat the analysis of the previous subsection setting now $\varepsilon = 0$ for all plots\footnote{Ideally speaking, DBI inflation corresponds to a vanishing coefficient $\alpha$~. Setting $\varepsilon = 0$ then sends the speed of sound to zero, or equivalently, $f_{\rm NL}$ to infinity, ruling out the DBI models with vanishing $\varepsilon$. This can be seen e.g. on the upper right panel of Fig. \ref{fig:3}, where the non-Gaussianity exclusion band covers the whole $\alpha=0$ axis. Of course, in a more realistic situation with a small but non-zero $\varepsilon$, an allowed region with $\alpha=0$ and a small-enough $\beta$ opens up.} but that corresponding to slow-roll WBG inflation (in which the regime with large non-Gaussianity crucially depends on the presence of a non-zero slow-roll parameter $\varepsilon$, see Eq. \eqref{srwbgngcond}.) In the latter case we set $\varepsilon = 10^{-3}$~.
Apart from these modifications, the figures \ref{fig:3} and \ref{fig:4} correspond to the exact same choices of parameters as in Figs. \ref{fig:1} and \ref{fig:2} respectively. 

The situation for $\epsilon \ll |n_s-1|$ is qualitatively similar to the previous case ($\epsilon\sim |n_s-1|$.) The current data still allows parameter space, consistent with the existing bounds on $r$ and corresponding to measurable non-Gaussianity. However, improving the limits on the amplitude of primordial gravitational waves could still induce dramatic changes. One novelty compared to the case of the previous subsection is that for $\epsilon \sim 0$, there would still remain a small allowed parameter space for DBI/G-/Galileon inflation even if $r < 10^{-2}$ (see the upper right panel of Fig. \ref{fig:4}). Moreover, a much larger fraction of the allowed parameter space would survive the $r < 10^{-2}$ bound in the slow roll WBG model, as seen from the lower right panels of Figs. \ref{fig:3} and \ref{fig:4}.

\section{Discussion and future directions}
\label{sec4}

In this paper, we have explored a number of constraints---both theoretical and experimental---on alternative theories to canonical slow-roll inflation, described by the general unitary-gauge action \eqref{action}. The latter action captures the great majority of single-field models motivated by considerations of theoretical consistency (in particular, stability under quantum corrections) and phenomenological interest (possibility to generate non-Gaussianity, detectable by current or near-future experiments.) The list of models covered by our analysis \eqref{action} includes, but is not limited to, DBI inflation \cite{Silverstein:2003hf, Alishahiha:2004eh}, G-inflation \cite{Kobayashi:2010cm}, Galileon inflation \cite{Burrage:2010cu}, the recently-proposed slow-roll inflation with weakly broken galileon symmetry \cite{Pirtskhalava:2015zwa}, etc. 

We have found that the requirement of the absence of instabilities and/or superluminal propagation, the current experimental limits on the tensor-to-scalar ratio and on non-Gaussianity result in an interesting interplay, ruling out complementary parts of the parameter space (see Figs. \ref{fig:1} through \ref{fig:4}.) The existing upper bound on the amplitude of primordial gravitational waves, $r\lsim 0.07$, still allows for an appreciable range of parameters for most of the above models, leaving room for detectable non-Gaussianity. The only theory which is already ruled out is the KWBG model \ref{kwbgpar}, characterized by an abrupt growth of non-Gaussianity, $f_{\rm NL}\propto 1/c_s^6$~, in the subluminal limit (note, that this model is excluded in a quite non-trivial way, however -- see the lower left panel of Fig \ref{fig:1}.) 

Reducing the upper bound on the tensor-to-scalar ratio down to $r< 10^{-2}$ would put all of the above-considered theories in some tension with experiment. This situation is depicted on Figs. \ref{fig:2} and \ref{fig:4}. The former figure corresponds to models with $\varepsilon\sim |n_s-1|$, in which case the only allowed region that has a chance of surviving our constraints (shown on the lower right panel) belongs to slow-roll inflation with weakly broken galileon symmetry of Ref. \cite{Pirtskhalava:2015zwa}. Fig. \ref{fig:4}, on the other hand, shows that the theories characterized by $\varepsilon\ll |n_s-1|$ have more chance of being consistent with small values of the tensor-to-scalar ratio. The SRWBG model seems to work slightly better compared to the other models also in this case, retaining a bigger fraction of its allowed parameter space with the decrease of the upper bound on $r$.
We thus find that, in addition to models driven by plateau-like potentials famously consistent with \eqref{smallr}, there would still remain a chance for an alternative single-field model driven by a \textit{convex potential} ($\varepsilon \sim |n_s-1|$), even if $r$ is measured to be as small as in Eq. \eqref{smallr}. While the latter model is slow-roll at the level of the background evolution, there are quantitative differences from the standard models with a canonical kinetic term. Most dramatically, the dynamics of perturbations is rather different in this model, allowing for somewhat strongly coupled and highly non-Gaussian scalar perturbations ($f_{\rm NL}\sim 1-20$) \cite{Pirtskhalava:2015zwa}.

Our analysis is rather general but it is not without loopholes, of course. We have argued above that, at the quadratic order, \eqref{action} is the most general action that captures theories characterized by scalar perturbations with usual, phonon-like dispersion relations \eqref{phdr}.
However, there exist models such as \textit{ghost inflation} \cite{ArkaniHamed:2003uz,Senatore:2004rj,Ivanov:2014yla}, where $\alpha\sim0$ and the background describes a perfect de Sitter space (i.e. $\varepsilon=0$), so that the scalar speed of sound vanishes at the zeroth order (consistent with our general expression for $c_s^2$, Eq. \eqref{cssq}.) In such a case, one ought to consider effects of higher-derivative operators in \eqref{action}---e.g. $\delta K^2$---that will dominate the gradient energy of the scalar modes at horizon crossing (i.e., at characteristic frequencies of order $\omega \sim H$.) This results in a rather different, $\omega = k^2/M$ dispersion relation with $M$ some cutoff scale, which would presumably modify our analysis. It would be interesting to see how our conclusions are affected in this class of models. 

While our theoretical calculation is complete, a precise analysis of the data on non-Gaussianity has remained outside of the scope of this work. The reason is that while the shape of non-Gaussianity generated by the effective theory \eqref{action} is always close to equilateral, it of course does not generically coincide with the templates constrained by experiments; therefore, extracting precision experimental limits is not straightforward. One possible improvement of our results (which we believe won't bring significant changes to our conclusions) would precisely be to do a precision analysis of the existing data on the scalar bispectrum in the context of our approach. 

Most importantly, however, our analysis reveals an interesting interplay between various theoretical and observational constraints that exclude complementary parts of the parameter space in general inflationary effective theories like the one of Eq. \eqref{action}. We believe that further developing this approach via incorporating more general effective theories and/or more precise handling of the data could be a useful step in better understanding the ultimate physics behind inflation.  

\subsection*{Acknowledgements}
It is a pleasure to thank Paolo Creminelli, Gregory Gabadadze, Riccardo Rattazzi and Sergey Sibiryakov for valuable discussions and comments. The work of E.T. is supported in part by MIUR-FIRB grant RBFR12H1MW.

\appendix
\section{The cubic  $\zeta$ action} 
\label{appa}

In this appendix, we derive the cubic Lagrangian for the comoving curvature perturbation $\zeta$ for a general effective theory described by the action \eqref{action}. The presentation closely follows and generalizes Maldacena's calculation of an analogous Lagrangian for canonical slow-roll inflation \cite{Maldacena:2002vr}, see also Ref. \cite{Chen:2006nt} for a generalization.
We will work with the ADM variables, $$ds^2=-N^2 dt^2 +g_{ij}(N^i dt+dx^i)(N^j dt+dx^j)$$ and denote by $\upleft{3}{R}$ the three-dimensional curvature corresponding to the metric $g_{ij}$ induced on equal-time hypersurfaces, while the tensor $E_{ij}$ is related to their extrinsic curvature $K_{ij}$ as follows
\beq
E_{ij} = N K_{ij} = \frac{1}{2}\(\p_t g_{ij}-\nabla_i N_j-\nabla_j N_i\)~.
\eeq
For a given inflationary spacetime, the first line of \eqref{action} is completely fixed by the background dynamics (i.e. by the Hubble rate and its first time derivative), while the second line (which starts at least quadratic in metric perturbations) is \textit{a priori} unconstrained. For a concrete inflationary model, the coefficients $M_2^4$, $M_3^4$, etc. will depend on the parameters of the underlying theory. 
A unitary-gauge action of the form \eqref{action} propagates a single scalar mode in addition to the usual transverse-traceless graviton. We exclusively concentrate on the dynamics of the former mode in this appendix, and ignore the tensor and the (non-dynamical) vector degrees of freedom from now on. The scalar mode is conveniently studied in the gauge defined by $g_{ij}= a^2(t) e^{2\zeta} \delta_{ij}$, where $\zeta$ has the meaning of the curvature perturbation of spatial hypersurfaces. In order to compute the cubic action for $\zeta$ in the theory \eqref{action}, we proceed as follows. We first integrate out the perturbations of the lapse and shift variables, $\delta N$ and $N_i\equiv \p_i\psi$, from their respective equations of motion, and then plug the solutions back into \eqref{action} to find the action for $\zeta$ at the desired order. To determine the cubic action, we only need to solve for the lapse and shift to the \textit{linear}  order  in $\zeta$ (see Ref. \cite{Chen:2006nt} for a discussion on this point.) The Hamiltonian and momentum constraint equations, obtained by varying the action w.r.t. $N$ and $N^j$, are
\begin{align}
\dfrac{M_P^2}{2} & \left[ \upleft{3}{R} -\frac{1}{N^2}\left( E^{ij}E_{ij}-E^2 \right) 
		+\dfrac{2}{N^2}\dot{H}
		- 2(3H^2+\dot{H})\right] 
	+ M_2^4\delta N 
	-\hat{M}^3_1 \delta E  =0 ,
\label{mg-Constraint-001}
\end{align}
\begin{equation}
\hat{\nabla}_i \left[ \frac{M_P^2}{N}\left(E^i_j - \delta^i_j E \right) 
	- \hat{M}^3_1 \delta^i_j \delta N \right]  = 0~,
\label{mg-Constraint-002}
\end{equation}
and one can readily find the linear solution
\begin{align}
\label{delta N}
\delta N& = \frac{2M_P^2}{2M_P^2H-\hat{M}^3_1} \dot \zeta \\
\label{Ni}
\psi &=-\frac{2M_P^2}{2M_P^2H-\hat{M}^3_1} \zeta + \frac{2M_P^2M_2^4-12M_P^2H\hat{M}^3_1+3\hat{M}^6_1-4M_P^4\dot{H}}{(2M_P^2H-\hat{M}^3_1)^2} \(\frac{\p}{a}\)^{-2}\dot\zeta\equiv C\zeta+\chi~.
\end{align}
When plugged back into the original action \eqref{action}, this yields the following quadratic Lagrangian for the comoving curvature perturbation
\begin{equation}
\begin{split}
S^{(2)} 	= \int\D^4x \, a^3 \mathcal{N} \bigg[
	\dot{\zeta}^2 - c_s^2\frac{(\partial_i\zeta)^2}{a^2}\bigg] ,
\label{mg-a2}
\end{split}
\end{equation}
where we have defined
\begin{align}
\mathcal{N} & \equiv M_P^2 \frac{-4M_P^4\dot{H}+2M_P^2M_2^4-12M_P^2H\hat{M}^3_1+3\hat{M}_1^6}{(2M_P^2H-\hat{M}^3_1)^2}~,
\label{mg-N}\\
c_s^2 & = \frac{4M_P^4\dot{H}-2M_P^2H\hat{M}^3_1+\hat{M}_1^6-2M_P^2\partial_t\hat{M}^3_1}{4M_P^4\dot{H}-2M_P^2M_2^4+12M_P^2H\hat{M}^3_1-3\hat{M}_1^6} ~.
\label{mg-cs}
\end{align}
Obtaining the cubic-order action requires significantly more work. A straightforward expansion of \eqref{action} using the solutions \eqref{delta N} and \eqref{Ni} for the lapse and shift perturbations, yields the following result 
\begin{equation}
\begin{split}
S^{(3)} &= \int\D^4x \, \bigg\{
	-a\mathcal{N}c_s^2\zeta(\partial_i\zeta)^2  
	+ a^3
	\left[C \mathcal{N}+\frac{C^3}{\Mpl^4}\lambda\right]\dot{\zeta}^3
	+ 3a^3\mathcal{N}\zeta\dot{\zeta}^2  +
\\
&	+ \frac{M_P^2}{2 a}\left(3\zeta+C\dot{\zeta}\right)
	\left[(\partial_i\partial_j\psi)^2-(\partial^2\psi)^2\right]
	- \frac{2M_P^2}{a}\partial_i\zeta\partial_i\psi\partial^2\psi +
\\
&	+ aC^2\(2\hat{M}_1^3-\hat{M}_2^3\) \dot{\zeta}^2\partial^2\psi
\bigg\}~,
\label{mg-a3}
\end{split}
\end{equation}
where
\begin{equation}
\lambda \equiv - \frac{M_P^2}{2}\left[2M_P^2(M_2^4+M_3^4)-3(2M_P^2H-\hat{M}_1^3)(2\hat{M}_1^3-\hat{M}_2^3)\right] .
\label{mg-lambda}
\end{equation}
It will prove convenient to recast the action \eqref{mg-a3} into a slightly different form. In doing so, we omit a number tedious but straightforward manipulations. Once the dust settles, one finds the following expression, equivalent to \eqref{mg-a3} up to a total derivative
\begin{equation}
\begin{split}
S^{(3)} 
	&= \int\D^4x \,  \bigg\{
	a^3
	\left[\mathcal{N}C\left(1+\frac{HC}{c_s^2}\right)
	- \lambda'
	\right]\dot{\zeta}^3 +
 \frac{a^3\mathcal{N}(HC)^2}{c_s^2}\left(\varrho-3+\frac{3c_s^2}{(HC)^2}\right) \zeta\dot{\zeta}^2 
\\
&	+ a\mathcal{N}(HC)^2\left(\varrho-2s+1-\frac{c_s^2}{(HC)^2}\right) \zeta(\partial_i\zeta)^2
	+ 2a\mathcal{N}HC\dot{\zeta}\partial_i\zeta\partial_i\chi 
	+\frac{a^3\mathcal{N}}{2}\frac{\D}{\D t}\left(\frac{n(HC)^2}{c_s^2}\right)\zeta^2\dot{\zeta} \\
&
	+ \frac{\mathcal{N}c_s^2}{2a}\partial_i\zeta\partial_i\chi\partial^2\chi +
	 \frac{\mathcal{N}c_s^2+2\hat{M}_1^3C}{4a}\partial^2\zeta(\partial_i\chi)^2 
	+ \frac{\hat{M}_1^3C^2}{a}\partial^2\zeta\partial_i\zeta\partial_i\chi
	+ \frac{\hat{M}_1^3C^3}{2a}(\partial_i\zeta)^2\partial^2\zeta \\&
	+ aC^3 (2\hat{M}_1^3-\hat{M}_2^3)\dot{\zeta}^2\partial^2\zeta 
	- g(\zeta)\frac{\delta L}{\delta\zeta}
\bigg\}~.
\label{our-14}
\end{split}
\end{equation}
In the last expression, we have defined a number of quantities 
\beq
n=\frac{1}{H}\frac{d}{dt}\log\(\mathcal{N} c_s^2\), \quad \varrho =\frac{1}{H}\frac{d}{dt}\log\(H C^2\), \quad s = \frac{1}{H}\frac{d}{dt}\log( c_s)~,
\eeq
as well as 
\begin{equation}
\lambda' \equiv C^3(M_2^4+M_3^4)
	+ C^2 (2\hat{M}_1^3-\hat{M}_2^3)\left(3-\frac{\mathcal{N}}{M_P^2}\right) .
\label{our-15}
\end{equation}
Furthermore, $\delta L/\delta \zeta$ denotes the variation of the quadratic action by $\zeta$,
\beq
\frac{\delta L}{\delta \zeta} = -2 \mpl^2 \p_t (a\p^2\chi) +2 a\mathcal{N} c_s^2 \p^2 \zeta~,
\eeq
and the coefficient of this term in the action \eqref{our-14} is
\begin{equation}
\begin{split}
g(\zeta) 
&	= \frac{n(HC)^2}{4c_s^2}\zeta^2+\frac{HC^2}{c_s^2}\zeta\dot{\zeta}
	+ \frac{C^2}{4a^2}\left[-(\partial_i\zeta)^2+\partial^{-2}(\partial_i\partial_j(\partial_i\zeta\partial_j\zeta))\right] +
\\
&\qquad
	- \frac{C}{2a^2}\left[\partial_i\zeta\partial_i\chi - \partial^{-2}(\partial_i\partial_j(\partial_i\zeta\partial_j\chi))\right] .
\label{our-g}
\end{split}
\end{equation}

As a quick consistency check, we note that our Eq.\eqref{our-14} reduces to an analogous expression, Eq. (4.26) of Ref. \cite{Chen:2006nt}, with the following substitutions: $\hat{M}_1^3=\hat{M}_2^3=0$, $C\rightarrow-H^{-1}$, $\mathcal{N}c_s^2\rightarrow\varepsilon$, $n\rightarrow\eta\equiv (d \log\varepsilon/dt)/H$ and $\varrho\rightarrow\varepsilon$. In that case,
a small speed of sound implies large (equilateral) non-Gaussianity, $f^{\rm equil}_{NL}\sim 1/c_s^2$~. 
Our result generalizes the cubic action of \cite{Chen:2006nt} to the case of non-zero $\hat{M}_1^3$ and $\hat{M}_2^3$, which opens up qualitatively novel ways of generating large non-Gaussianity, as explained in the main text.

\section{Bispectrum}
\label{appb}

We follow the standard nomenclature in defining various quantities associated with the two- and three-point functions of the comoving curvature perturbation $\zeta$, see, e.g., Ref. \cite{Babich:2004gb}. 
The three-point function is defined as 
\beq
\langle \zeta(\vec k_1)\zeta(\vec k_2)\zeta(\vec k_3)\rangle=(2\pi)^3\delta^{(3)}\(\sum_i \vec{k}_i\)B_\zeta(k_1,k_2,k_3)~,
\eeq
 and $k_t\equiv k_1+k_2+k_3$~. We will at times use a related quantity, $B_{\Phi} = (3/5)^3 B_{\zeta}$~, for the Bardeen potential $\Phi=(3/5)\zeta$~. Moreover, we will denote 
\beq 
A =  H^2 \mathcal{N}~,
\eeq
where $\mathcal{N}$ is the normalization factor for the $\zeta$ kinetic term.
The scalar perturbations' power spectrum can be directly read off Eq. \eqref{mg-a2} 
\beq
P_\zeta (k) = \frac{1}{4} \frac{H^4}{A}\frac{1}{(c_sk)^3}~,
\eeq
and dimensionless power spectrum is
\beq
\label{dimlessps}
\Delta^2_\zeta =\frac{k^3}{2\pi^2}P_\zeta (k)\simeq 5\pi^2\cdot 10^{-9}~.
\eeq
With all of the above qualifications, the contribution from each of the relevant cubic operators (we only exclude the operators which are suppressed by at least two powers of slow-roll) in \eqref{our-14} to the bispectrum reads
\beq
B_{\Phi}(k_1,k_2,k_3)= 2 \(\frac{3}{20}\)^3 \frac{H^{10}}{A^3 c_s^6}\sum_i c_i S_i(k_1,k_2,k_3) + \text{cyclic}~.
\eeq
\paragraph{1. Contribution from $\dot \zeta^3$}
\begin{gather}
S_1(k_1,k_2,k_3) =\frac{4}{k_t^3 k_1k_2k_3}  \\
c_1=-\frac{\mpl^2}{c_s^4}~\frac{\alpha^2-\alpha-\varepsilon+ c_s^2\(3\alpha-\gamma+1\)\(\alpha^2-\alpha-\varepsilon\)+c_s^4 (\alpha-1)\(6\alpha^2-3\alpha\gamma+3\gamma+\delta-2\varepsilon\)}{(\alpha-1)^4}
\end{gather}

\paragraph{2. Contribution from $\zeta \dot \zeta^2$}
\begin{gather}
S_2(k_1,k_2,k_3) = 2~\frac{k_t+k_1}{k_t^2 k_1^3 k_2k_3} \\
c_2= 3~\frac{\mpl^2}{c_s^4}~ \frac{(\alpha^2-\alpha-\varepsilon)\(1-(\alpha-1)^2 c_s^2\)}{(\alpha-1)^4}
\end{gather}
\paragraph{3. Contribution from $\zeta (\p \zeta)^2$}
\begin{gather}
S_3(k_1,k_2,k_3) = (k_1^2-k_2^2-k_3^2)\(-\frac{k_t}{(k_1k_2k_3)^3}+\frac{\sum_{i>j}k_ik_j}{k_t (k_1k_2k_3)^3}+\frac{1}{k_t^2(k_1k_2k_3)^2}\)\\
c_3 = -\frac{c_2}{3} 
\end{gather}
\paragraph{4. Contribution from $\dot\zeta\p_i\zeta\p_i\p^{-2}\dot\zeta$}
\begin{gather}
 S_4(k_1,k_2,k_3) =\frac{k_1^2-k_2^2-k_3^2}{2 k_t k_1(k_2k_3)^3}~\(2+\frac{k_2+k_3}{k_t}\) \\
 c_4 = -\frac{\mpl^2}{c_s^4}~\frac{(\alpha^2-5\alpha-\varepsilon+4)(\alpha^2-\alpha-\varepsilon)^2}{2(\alpha-1)^6}
\end{gather}
\paragraph{5. Contribution from $\p^2\zeta \p_i\p^{-2}\dot\zeta \p_i\p^{-2}\dot\zeta$}
\begin{gather}
  S_5(k_1,k_2,k_3) = \frac{k_1^2-k_2^2-k_3^2}{k_t k_1(k_2k_3)^3}~\(1+\frac{k_1}{k_t}\) \\
c_5 = \frac{\mpl^2}{c_s^4}~\frac{(\alpha^2-\alpha-\varepsilon)^2(3\alpha^2-3\alpha+\varepsilon)}{4(\alpha-1)^6}
\end{gather}
\paragraph{6. Contribution from $\p^2\zeta\p_i\zeta\p_i\p^{-2}\dot\zeta$}
\begin{gather}
 S_6(k_1,k_2,k_3) = \frac{k_1^2-k_2^2-k_3^2}{2 k_t k_1(k_2k_3)^3}~\(2+\frac{2 k_1+k_2+k_3}{k_t}+\frac{2k_1(k_2+k_3)}{k_t^2}\) \\
 c_8 = -\frac{\mpl^2}{c_s^4} ~\frac{2\alpha (\alpha^2-\alpha-\varepsilon)}{(\alpha-1)^4}
\end{gather}
\paragraph{7. Contribution from $\p^2\zeta(\p\zeta)^2$}
\begin{gather}
S_7(k_1,k_2,k_3) =2~ \frac{k_1^2-k_2^2-k_3^2}{k_t k_1(k_2k_3)^3}~\(1+\frac{\sum_{i>j}k_ik_j}{k^2_t}+\frac{3k_1k_2k_3}{k_t^3}\)\\
c_7 = \frac{\mpl^2}{c_s^4}~\frac{\alpha}{(\alpha-1)^3}
\end{gather}
\paragraph{8. Contribution from $\p^2\zeta\dot\zeta^2$}
\begin{gather}
S_8(k_1,k_2,k_3) =4~ \frac{k_t+3k_1}{k_t^4k_1k_2k_3}\\
c_8 = \frac{\mpl^2}{c_s^2}~\frac{4\alpha-\gamma}{(\alpha-1)^3}
\end{gather}
We have left out the operator $\zeta^2\dot\zeta$ in \eqref{our-14}, since its coefficient is at most of order $\sim \varepsilon ^2$, and it is not expected to affect our analysis in any significant way. Moreover, the last term in Eq. \eqref{our-14}, being proportional to the lower-order $\zeta$ equation of motion, can be removed by a field redefinition. The latter redefinition also contributes to the three-point function of the conserved scalar mode through the function $g(\zeta)$ in \eqref{our-g}. The contributions of the terms in this function that include derivatives are suppressed at superhorizon distances. Moreover, the first term contributes to $f_{\rm NL}$ by an amount that scales as $\eta/c_s^2$ (see, e.g. \cite{Chen:2006nt}). We neglect this piece in our analysis, since it is always expected to be sub-dominant (whenever non-trivial constraints from bispectrum arise, there are leading contributions, enhanced by at least a factor of $1/\eta$ compared to it.)

\section{The regime with $f_{\rm NL}\propto \frac{1}{c_s^6}$}
\label{appc}

In this appendix, we show how the behavior \eqref{largeng} arises in the model \eqref{kwbgpar}.
For our purposes, it will be sufficient to set $\varepsilon=0$ to avoid over-complication of expressions. 
In order to study the part of the parameter space, described by \eqref{kdwbg}, it will prove convenient to slightly change the notation by defining
\beq
x\equiv \frac{1-\alpha}{c_s^2}~,
\eeq
so that $x$ is generically an order-one constant. 
The normalization factor for the curvature perturbation then becomes
\beq
\mathcal{N} = \mpl^2 ~\frac{1-x c_s^2}{x c_s^4}~.
\eeq
The latter quantity has a strong dependence on the speed of sound: for small $c_s^2$, it grows like $\mathcal{N}\propto 1/c_s^4$, and this appears to make the scalar perturbations weakly coupled -- suppressing the self-interactions of the canonically normalized $\zeta$, and therefore suppressing non-Gaussianity. This observation is decisive, however: in order to make a conclusive statement regarding non-Gaussianity, one has to study the $c_s$-dependence of the cubic $\zeta$ interactions in the theory at hand. In fact, we will find that the cubic interactions grow as fast as $1/c_s^{10}$, eventually leading to non-Gaussianity of order $f_{\rm NL}\propto 1/c_s^6$. 
The action for the comoving curvature perturbation, up to the cubic order in non-linearity, is given in Eq. \eqref{our-14}. In the remainder of this section, we will confine ourselves to the leading order in the $1/c^2_s$-expansion, which allows to extract the fastest-growing effects in the deep subluminal region of the parameter space of interest. There are seven operators that contribute in this limit, and assuming $\alpha\simeq 1$, the relevant cubic action becomes
\ba
\label{reducedaction}
S^{(3)}_\zeta &=&\mpl^2 \int d^4 x ~a^3 ~ \bigg\{ \frac{1}{x c_s^4} \bigg[ \dot\zeta^2 -c_s^2 \frac{(\p\zeta)^2}{a^2}  \bigg]  +\frac{1}{x^3 c_s^{10}}\bigg[\frac{1}{H}\dot\zeta^3 - 3\zeta\dot\zeta^2   \nonumber + c_s^2 \zeta \frac{(\p\zeta)^2}{a^2}  \nonumber \\&-&\frac{3}{2} \dot\zeta\p_i\zeta\p_i\p^{-2}\dot\zeta   
-\frac{3}{4}\p^2\zeta \p_i\p^{-2}\dot\zeta \p_i\p^{-2}\dot\zeta   +2\frac{c_s^2}{H}\frac{\p^2\zeta}{a^2}\p_i\zeta\p_i\p^{-2}\dot\zeta -\frac{c_s^4}{H^2} \frac{\p^2\zeta}{a^2} \frac{(\p\zeta)^2}{a^2}  \bigg] \bigg\}~.
\ea
Various terms in \eqref{reducedaction} appear to be of different order in $1/c_s^2$; however, one should keep in mind that higher spatial derivatives lead to additional factors of $1/c_s$ in the amplitude of non-Gaussianity, so that e.g. the two cubic operators $\dot\zeta^3/c_s^{10}$ and $\zeta (\p\zeta)^2/ c_s^8$ contribute comparably to $f_{\rm NL}$ in the small $c_s^2$ limit. 

\begin{figure}
\centering
\includegraphics[width=.75\textwidth]{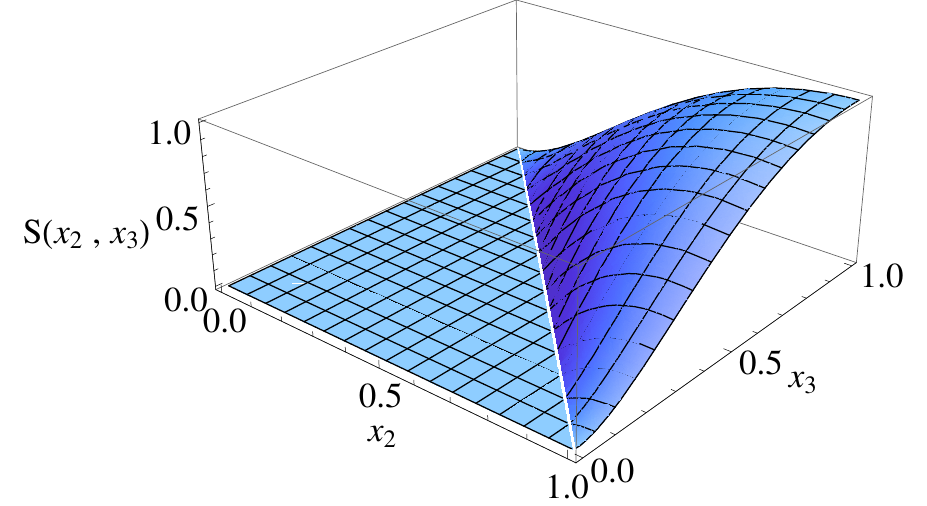} 
\caption{The shape of non-Gaussianity, corresponding to Eq. \eqref{limitnongauss} (we use the standard notation: $x_2 = k_2/k_1$ and $x_3 = k_3/k_1$.)}
\label{fig:shape}
\end{figure}

Summing up all the contributions to non-Gaussianity described in the previous appendix yields a simple expression at the leading order in the $1/c_s^2$ expansion
\ba
\label{limitnongauss}
B_\Phi\(k_1,k_2,k_3\)&=&-12 \(\frac{3}{20}\)^3\frac{\mpl^2 H^{10}}{(Ax)^3 c_s^{16}}\frac{1}{k_t^3(k_1k_2k_3)^2} \bigg [\sum_i k_i^3 -\sum_{i\neq j} k_i^2k_j+2k_1k_2k_3\bigg ]~.
\ea
The shape of the bispectrum is close to the equilateral one, see Fig. \ref{fig:shape}. We note, that the second and the third operators above lead to squeezed non-Gaussianity, which can not characterize a derivatively coupled theory of the sort we are considering; and indeed, all `non-equilateralness' cancels out in the full bispectrum, which is a nice consistency check of our results. This leaves us with the simple expression in Eq. \eqref{limitnongauss}. 

The fact that all of these terms are equally important at typical frequency scales of order of the inflationary Hubble rate can also be seen from the analysis of the $2\to2$ scattering of $\zeta$ quanta. 
It is a fairly generic fact, that the strong coupling scale of a theory shrinks to zero (or, more precisely, becomes dominated by higher-order effects in the dispersion relation) as the speed of sound is taken to vanish, and in the limit of small $c_s^2$ that we are interested in here, one should be extra careful with this scale. In particular, in order to be able to trust our low-energy effective theory, we should make sure that the strong coupling scale of that theory is parametrically greater than the Hubble rate $H$ --- the typical energy/frequency scale that we measure the inflationary observables at. In order to assess the strong coupling scale $\Lambda_\star$ associated with the $i$-th cubic interaction in \eqref{reducedaction}, one can study a $2\to 2$ scattering of $\zeta$ that stems from that interaction. These interactions are non-relativistic, however, and extra care has to be taken in order to properly define the energy and momentum scales at which perturbative unitarity is violated in the scattering of interest.  
There are various ways of addressing these issues, and perhaps the shortest and the most straightforward one is via formally switching back to the relativistic notation. To this end, one can rescale time variable in the action \eqref{reducedaction}, so that $t= \tilde t/c_s$, and $\omega=\tilde \omega c_s$ for the frequency. The quadratic $\zeta$ action then becomes
\beq
S_\zeta^{(2)} = \mpl^2\int d^3x d\tilde t~ a^3~ \frac{1}{x c_s^3} ~\bigg[ \(\p_{\tilde t}\zeta\)^2 - \frac{(\p\zeta)^2}{a^2}  \bigg]~,
\eeq
while the first cubic interaction in \eqref{reducedaction} reads
\beq
S_\zeta^{(3)} \supset \mpl^2 \int d^3x d\tilde t~ a^3~\bigg[ \frac{1}{x^3 c_s^8 H}\(\p_{\tilde t}\zeta\)^3 +\dots  \bigg]~.
\eeq
Canonically normalizing the curvature perturbation, $\zeta = x^{1/2} c_s^{3/2}\zeta_c$~, one immediately obtains the strong coupling scale associated with the given operator $\tilde \Lambda_{\star}^2 = x^{3/2} c_s^{7/2} \mpl H$. Note that this scale is still defined in the rescaled temporal coordinates; going back to the original coordinates, one finally arrives at the true strong coupling (energy) scale
\beq
\Lambda^2_{\star} = c_s^2 \tilde \Lambda^2_{\star} = x^{3/2} c_s^{11/2} \mpl H~.
\eeq
Demanding that this scale be larger than the inflationary Hubble rate yields the following bound
\beq
\label{hubblecond}
 x^{3/2} c_s^{11/2} \mpl > H~.
\eeq
What about the rest of the cubic operators in \eqref{reducedaction}? We have seen, that they all contribute by an equal order of magnitude to non-Gaussianity, so that there should exist a well-defined sense in which they are all `equally strongly coupled' around the Hubble frequencies. 
Repeating the above analysis, it is easy to see that the six remaining cubic interactions in \eqref{reducedaction} imply strong coupling scales, in general different from $\Lambda_\star$. For example, the operator $\zeta \dot\zeta^2$ starts violating perturbative unitarity in a $2\to 2$ scattering of $\zeta$ quanta around the frequency scales of order
\beq
\bar\Lambda_{\star} = x^{3/2} c_s^{11/2} \mpl ~.
\eeq
Being different from $\Lambda_\star$, the expression for $\bar\Lambda_\star$ nevertheless implies the exact same condition \eqref{hubblecond} if the scattering at Hubble frequencies is to be unitary. One can straightforwardly check that the same conclusion applies to all operators in \eqref{hubblecond}, fixing the sense in which all of these operators are equally important for the physics at the horizon.

\bibliographystyle{utphys}
\addcontentsline{toc}{section}{References}
\bibliography{eftinf}

\end{document}